\documentclass[review]{elsarticle}

\usepackage{graphics}
\usepackage{epstopdf}
\usepackage{adjustbox}
\usepackage{hyperref}
\usepackage{amsfonts}
\usepackage{amsmath}
\graphicspath{ {./figures/} }
\usepackage{pifont}
\newcommand{\xmark}{\text{\ding{55}}}

\usepackage{lineno,hyperref}
\modulolinenumbers[5]

\journal{Journal of \LaTeX\ Templates}

\begin{document}

\begin{frontmatter}

\title{Shape related constraints aware generation of Mechanical Designs through Deep Convolutional GAN }

\author[email_expleo,email_ensam]{ALMASRI Waad}

\fntext[email_expleo]{
Email address: firstName.lastName@expleogroup.com \\ 
Address : Expleo France,\\ 
Montigny Le Bretonneux, 78180, France }

\fntext[email_ensam]{
Email address: firstName.lastName@ensam.eu \\ 
    Address : LISPEN Laboratory\\
    Ecole Nationale Supérieure des Arts et Métiers ENSAM,
    Chalon-Sur-Saône,  France}

\author[email_expleo]{BETTEBGHOR Dimitri }
\author[email_ensam]{ABABSA Fakhreddine}
\author[email_ensam]{DANGLADE Florence}

\begin{abstract}
Mechanical product engineering often must comply with manufacturing or geometric constraints related to the shaping process.  Mechanical design hence should rely on robust and fast tools to explore complex shapes, typically for design for additive manufacturing (DfAM). Topology optimization is such a powerful tool, yet integrating geometric constraints (shape-related) into it is hard. In this work, we leverage machine learning capability to handle complex geometric and spatial correlations to integrate into the mechanical design process geometry-related constraints at the conceptual level. More precisely, we explore the generative capabilities of recent Deep Learning architectures to enhance mechanical designs, typically for additive manufacturing. 
 In this work, we build a generative Deep-Learning-based approach of topology optimization integrating mechanical conditions in addition to one typical manufacturing condition (the complexity of a design i.e. a geometrical condition). The approach is a dual-discriminator GAN: a generator that takes as input the mechanical and geometrical conditions and outputs a 2D structure and two discriminators, one to ensure that the generated structure follows the mechanical constraints and the other to assess the geometrical constraint.
We also explore the generation of designs with a non-uniform material distribution and show promising results. Finally, We evaluate the generated designs with an objective evaluation of all wanted aspects: the mechanical as well as the geometrical constraints.
\end{abstract}

\end{frontmatter}

\section*{INTRODUCTION}

\begin{figure}
\begin{center}
\includegraphics[width=16cm]{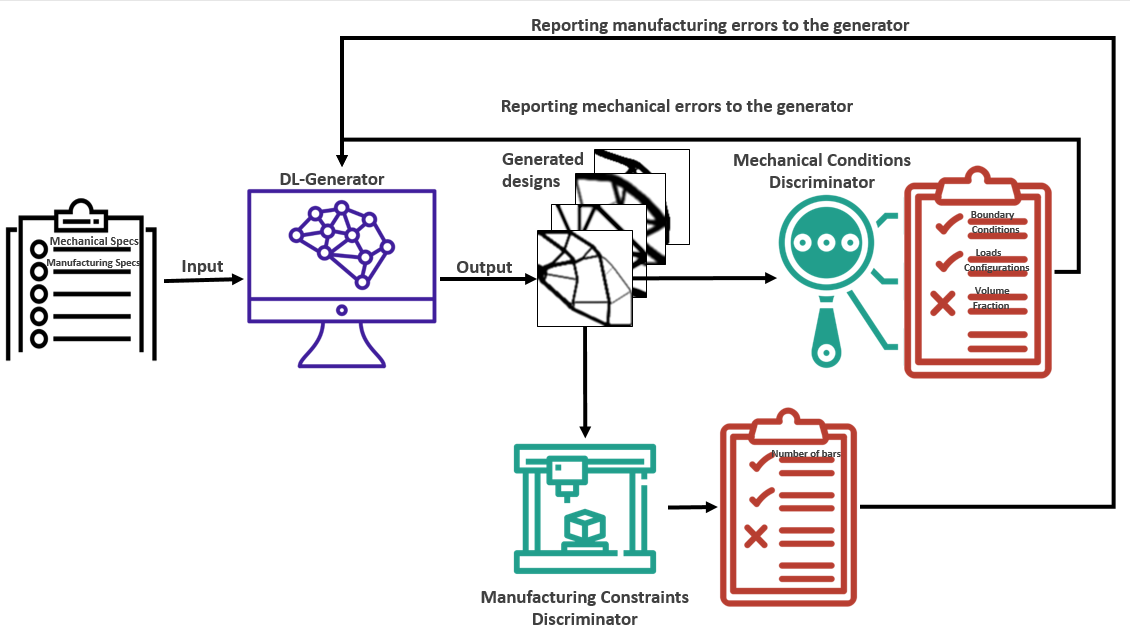}
\caption{GAN-BASED APPROACH FOR MANUFACTURABLE DESIGN GENERATION. The mechanical and manufacturing constraints are input into the generator. The latter outputs the corresponding designs which are fed afterwards into different types of validators: the mechanical discriminator which penalizes the generator when the generated designs do not comply with the mechanical conditions, and the manufacturing discriminator which penalizes the generator when the generated designs do not comply with the manufacturing conditions.} 
\end{center}
\label{figure_abstract_sktech} 
\end{figure}

The development of ultimately complex and efficient products requires complex mechanical shapes to be manufactured. The advent of new shaping processes, such as additive manufacturing, makes it possible to produce shapes that were previously unthinkable with conventional shaping processes. Although these new processes have their own manufacturing or geometric constraints, they do possess a great potential to improve product value. 
In the latest of the \(20^{th}\) century, additive manufacturing (AM) surfaced to the industrial world, pulling with it topology optimization into becoming one of the most hot research topic. AM facilitated the transmission of complex designs from paper into real mechanical and structural parts\cite{adam2014design}. Despite the success of AM, not all the designs could be manufactured: steep curvatures, sharp corners, overhanging patterns, the need for supports or scaffolds and other geometrical constraints are still a hurdle\cite{adam2014design}. And that is what motivated researchers to integrate geometrical constraints into topology optimization to have a better control over the shape and hence the manufacturability of the parts. 
Leary et al.(2014)\cite{leary2014optimal} proposed a method that alters the optimal topology output by SIMP to allow an additive manufacture without the use of support materials.
Mass and Amir (2017)\cite{mass2017topology} developed a topology optimization (TO) approach that accounts for overhang limitations. Zhang and Zhou (2018)\cite{zhang2018topology} adapted the topology optimization method not only to consider overhang constraints but also to produce self-supporting structures that can be directly printed by AM. 
Li et al.(2020)\cite{li2020additive}  proposed an AM-driven topology optimization method, in which he proved experimentally that considering the building direction in the topology optimization improved the mechanical performance of the printed designs when compared to randomly oriented ones. Wang and Qian (2020)\cite{wang2020simultaneous} suggested a density-based approach that optimizes the building direction and topology of a structure simultaneously while reducing the internal and external supports.
In this paper, we take a different perspective. The primary objective is to integrate geometrical constraints to the design phase. However, such constraints are spatial-related and cannot be easily formulated using mechanical-based equations. Thus, we proceed using image recognition techniques including among others the Deep Learning (DL) to learn these geometrical constraints alongside the mechanical ones in the design phase. In other terms, we seek to explore DL techniques in generating mechanical design and their potential of learning to reproduce and control the wanted geometrical constraint simultaneously in order to have a final manufacturable design. In this study, the designs being dealt with are truss-like structures, thus, a typical geometrical manufacturability constraint is chosen, the complexity i.e. the total number of bars in a design (Fig. 1\ref{figure_abstract_sktech}).
Among the various DL techniques, the generative methods have gained great success recently for their ability to learn the real distribution of the data and perhaps accelerate and improve the generation of different types of data. The top most common methods are the variational auto-encoder (VAE)~\cite{kingma2013auto} and the generative adversarial networks (GAN)~\cite{goodfellow2014generative}. VAE learns to generate dispersed samples via the minimization of a pixel-wise reconstruction error between the generated and input samples, making the VAE stable during training. However, this stable training, which consists of optimizing the average reconstruction loss, suffers from a major drawback: blurry generated images. In 2014, Goodfellow introduced the GAN architecture\cite{goodfellow2014generative}. It outperformed VAE in generating sharper, aesthetically more plausible and more creative images.
Nonetheless, GAN suffers from many issues during training with the most common being: difficulty in balancing the learning speed between the generator and the discriminator resulting in a mode collapse (i.e. when the generator learns to only produce one mode of samples) and an instability of the  generator and discriminator's losses (they oscillate continuously and never converge into a fixed point). Currently, research is focusing on solving these issues: Arjovsky et al. (2017)\cite{arjovsky2017wasserstein} proposed to replace the Kullback-Leibler loss function by the earth mover distance. Metz et al. (2016)\cite{metz2016unrolled} used an unrolled optimization of the discriminator's objective during training. Others explored different training hacks (batch normalization, transposed convolutions, the choice of activation functions and optimizers, etc.)\cite{radford2015unsupervised} etc. 
In this work, the objective is not only to generate designs aesthetically valid but also to evaluate them over mechanical and geometrical constraints, in other terms integrate mechanical and geometrical discriminators. Consequently, GAN is a better fit in this case, for its flexibility in adding discrimination blocks, which is not straightforward in VAE.

The contributions of this paper are summarized as follow:
\begin{itemize}
  \item With this approach, we can control the geometrical constraint,
  \item and we integrate non uniform material distribution into the topology optimization process, which is much complicated with its traditional formulation.
  \item The generator can be generalized to any other 2D designs output by other procedures than topology optimization.
  \item We formulate an objective evaluation to assess the generated designs and evaluate their conformity to the different imposed constraints.
  \item The traditional and bar counter discriminators, as well as the regression compliance predictor can be used separately and integrated in other works. These modules can also be replaced by other types of modules depending on the information we want to incorporate into the generation procedure such as identifying the structures needing higher manufacturing time through a build time discriminator, learning to reproduce the non-linear material, or geometrical effects in the design through a non-linear effects discriminator or the thermal distortion and other complex modules necessary to validate a design mechanically and geometrically.
\end{itemize}
The paper is organized as follows: section~\ref{section:related_works} details the state-of-the-art of integrating artificial intelligence into the topology optimization process, sections~\ref{section:Topo} and~\ref{section:gan} provide a theoretical overview of the topology optimization and the generative adversarial networks respectively. In section~\ref{section:method}, the novel GAN-based approach is detailed: the generator/discriminators' architectures, the parameters, the loss function, etc. Section~\ref{section:compliance} presents the Deep-Learning-based compliance predictor. Section~\ref{section:results} details the consolidation of the datasets used to train, validate and test our GAN-approach, inspects the generated designs and evaluates its potential into learning the mechanical and geometrical constraints. And finally, section~\ref{section:conclusion} summarizes the methodology and its outcomes and discusses future works.

\section{Related Works}\label{section:related_works}

\begin{table}[hbt!]
\caption{OUR APPROACH VERSUS THE STATE-OF-THE-ART: EVALUATION METRICS AND CONTRIBUTIONS. In this table, we compare our evaluation to the ones used in previous works and show our contributions.}
\label{table_evalutation}
\begin{center}
{\small
\begin{tabular}{|p{2cm}|p{3.2cm}|p{1.2cm}|p{1.5cm}|p{1.2cm}|p{2.2cm}|p{1.5cm}|}
\hline Author & Aesthetics & Volume Fraction & Compliance & Time Speedup & Non-Uniform Volume Fraction & Complexity \\ 
\hline Ulu et al. (2016)\cite{ulu2016data} & L1 distance (MAE) & \xmark  & \checkmark & \checkmark  & \xmark  & \xmark \\
\hline Oh et al. (2018)\cite{oh2018design} & subjective evaluation & \xmark & \xmark & \xmark & \xmark  & \xmark \\ 
\hline Sosnovik et al. (2019)\cite{sosnovik2019neural} & Binary Accuracy (Bin Acc), Intersection over Union (IoU) & \xmark& \xmark & \xmark & \xmark & \xmark \\
\hline Yu et al. (2019)\cite{yu2019deep} & MAE & \checkmark & \checkmark & \checkmark & \xmark  & \xmark \\
\hline Sharpe and Seepersad (2019)\cite{sharpe2019topology} & subjective evaluation & \checkmark & \xmark & \xmark & \xmark & \xmark \\
\hline Rawat and M-H Herman (2019)\cite{rawat2019novel} & subjective evaluation & \checkmark & \xmark & \checkmark & \xmark  & \xmark  \\
\hline Abueidda et al. (2020)\cite{abueidda2020topology} & Dice similarity coefficient (DSC) & \xmark & \xmark & \xmark & \xmark  & \xmark \\
\hline Nie et al. (2020)\cite{nie2020topologygan} & MAE, MSE (Average Reconstruction error) & \checkmark & \checkmark & \xmark & \xmark  & \xmark  \\
\hline Malviya (2020)\cite{malviya2020systematic} & subjective evaluation, MAE & \xmark & \xmark & \xmark & passive/active elements  & \xmark \\
\hline Our approach & subjective evaluation, MSE & \checkmark & \checkmark & \checkmark & \checkmark & \checkmark \\
\hline
\end{tabular}
}
\end{center}
\end{table}
With the emergence of 3D printing processes, mainly additive manufacturing, printing designs of higher complexity and creativity became doable\cite{adam2014design}, and thus made topology optimization further attractive in the research areas. Topology optimization aims at finding the optimal design layout given a set of parameters: loads, boundary conditions and other constraints. It finds the optimal material distribution inside a design domain such that the obtained structure has optimal mechanical properties and satisfies the prescribed constraints. The resulting design can have any shape (with curves and fine details) which can be defined as a 2D binary image or a 3D binary voxel grid such that the presence of a non-empty pixel/voxel means the presence of material. Topology optimization gained great success in the \(20^{th}\) century during the industrial revolution of the automotive and aerospace domains, given its powerful potential of optimizing a structure in terms of material used, while maintaining its recommended mechanical specifications and properties. Afterwards, it has spread its applications to a wider range of disciplines: fluids, acoustics, electromagnetics, optics, etc. However, topology optimization uses an iterative, finite-element-based method that is computationally expensive and hence relatively slow. 

In fact, with the new advanced software in topology optimization, outputting a single design subjected to a set of boundary conditions and load configurations is a matter of seconds to hours depending on the complexity of the design and its surrounding constraints. This is acceptable if this process is limited to this single step. However, a mechanical engineer never relies on the first design he/she tries, especially since topology optimization, in its commercially available form, does not consider complex mechanical criterion (such as non linear material or non linear geometric effect) nor manufacturing criteria nor other customized industrial constraints (automotive, aeronautic, hydraulic, etc.). The mechanical engineer has always to explore several set-ups in order to find the optimal topology, shape and sizing of the design to be manufactured. He/She has also to ensure that its final draft is creative, cost efficient and manufacturable. Hence, this accumulated iterative exploration process can become slow especially for large-scale designs.
Consequently, researchers try to find workarounds to bypass these constraints and accelerate the design optimization phase using DL techniques.
On the other hand, the emergence of DL showed promising results in many supervised and unsupervised tasks: from 2D object segmentation \cite{he2016deep, szegedy2015going, ioffe2015batch, szegedy2016rethinking, szegedy2017inception} to 3D complex object generation \cite{kar2015category, choy20163d, wu20153d}. 
Among the various DL techniques, the generative methods have dominated recently for their ability to learn the real distribution of the data and perhaps accelerate and improve the generation of different types of data. 
Particularly, GAN showed robustness and reliability in generating realistic face images \cite{karras2017progressive}, increasing the resolution of images while preserving their high frequency details \cite{ledig2017photo}, and generating 3D designs of several categories (cars, airplanes, bicycles, etc.) \cite{wu2016learning, zhu2018learning}.
Consequently, some researchers have tried to accelerate the topology optimization process via machine and DL techniques. \\
Ulu et al. (2016)\cite{ulu2016data} used Principal Component Analysis (PCA) to encode high-dimensional input configuration (boundary conditions, load configurations, etc.) to lower dimensional features which were input into a shallow neural network to output the final optimal structure. Oh et al. (2018)\cite{oh2018design} used a Boundary Equilibrium Generative Adversarial Network (BEGAN) to generate creative and aesthetically plausible wheel designs followed by a topology optimization code to assure mechanical validity. Sosnovik et al. (2019)\cite{sosnovik2019neural} formulated the topological problem as an image segmentation task and built an auto-encoder, which maps an intermediate structure outputted by traditional topology optimization solver to its final structure, thus always depending on the iterative FEM method. Kallioras et al. (2020)\cite{kallioras2020accelerated} followed the same procedure and integrated a Deep Belief Network between the first and last phases of SIMP in order to accelerate the optimization process. The following researchers took another perspective. Yu et al. (2019)\cite{yu2019deep} learned the mapping from boundary conditions to a high-resolution optimal structure using two steps: an auto encoder that encodes the boundary constraints and then outputs a low resolution design followed by a a conditional GAN that reconstructs its original resolution. On the other side, Sharpe and Seepersad (2019)\cite{sharpe2019topology} used directly a conditional GAN in order to output the optimal structure given the boundary conditions as compact vectors in order to explore further designs (with different sets of constraints) in a shorter time. Simultaneously Rawat and M-H Herman (2019)\cite{rawat2019novel} explored a conditional Wasserstein GAN to generate 2D designs given the volume fraction as the input condition. However, they encountered a mode collapse: the generator succeeded to output only valid designs for one value of the volume fraction condition (0.4). Abueidda et al. (2020)\cite{abueidda2020topology} used the ResUnet architecture in order to generate 2D structures with nonlinearities; the encoder compresses high dimensional boundary conditions, loads configurations, volume fractions, etc. into low-dimensional matrices and the decoder reconstructs the corresponding optimal 2D structure. Nie et al. (2020)\cite{nie2020topologygan} generated 2D structures using a conditional Pixel2Pixel GAN architecture: the boundary conditions are transformed into physical fields like the von Mises stress fields, strain energy fields and displacement fields, the latter are inputted to a U-SE-ResNet\footnote{SE stands for Squeeze excitation\cite{Hu_2018_CVPR}} generator to output the optimal designs. The discriminator takes the boundary conditions, the physical fields all along with the designs and outputs the probability of the design being real. In their work, Nie et al. still relied on a FE computation to transform the mechanical conditions into physical numerical fields. Malvia (2020)\cite{malviya2020systematic} tested the performance of four different types of CNN and GAN-based networks in generating 2D designs given the boundary, loads and volume fraction conditions in the objective of exploring the potential of deep learning techniques and data augmentation in accelerating the process of TO. Particularly, Malvia\cite{malviya2020systematic} trained their networks on volume fraction matrices with passive and active elements. \\ \\
In this research, we complement the work done before by taking into consideration not only boundary conditions but also the complexity of the design. As we know, an industrial design is characterized by not only its mechanical specifications but also by its creativity, feasibility and manufacturability. On one hand, not all designs outputted by topology optimization can be manufactured for several reasons, among them the complexity. On the other hand, the manufacturability constraints are not yet standardized and included in the topology optimization process. In this work, we consider one manufacturability constraint, the complexity which is defined as the number of bars forming a design and will be referred to by the term "geometrical condition".
In fact, many industrial topology optimization software like Tosca Abaqus\cite{tosca}, Inspire\cite{inspire} of solid Thinking, Within\cite{within} / NetFabb\cite{netfabb} of Autodesk, ProTOp\cite{protop} of Caess etc. integrate some mechanical conditions along with the geometrical ones as inputs, then run a finite-element-based computation to finally output the final optimal structure. In this work, we use a full DL generative pipeline that reproduces the topology optimization process without having to run an iterative algorithm and that is able to generate designs faster, thus, enabling a fast and easy exploration of designs by simply setting the boundary and geometrical conditions. Such a fast generative process can also be useful to output more robust designs when load cases and boundary conditions feature uncertainty. One major originality of our work is that our approach allows to consider non uniform density distribution (or what we will be referring to non-uniform volume fraction). Therefore, we explore the generation of designs with non-uniform volume fraction and designs with passive/active elements. \\
Unlike most of the work done using generative methods, and which consisted of judging subjectively the aesthetics of the generated designs, in this paper, we evaluate the generation ``objectively'' by computing three metrics: the mechanical validity (here compliance), the volume fraction i.e. the percentage of material used and the complexity of the generated designs; the compliance and complexity evaluators are DL-based networks. Table \ref{table_evalutation} summarizes the contribution of this paper in comparison with the state of the art. \\
This work is inspired from the Generative adversarial network approach\cite{goodfellow2014generative}. It is a dual-discriminator GAN. The first discriminator identifies whether an input design is real (design generated by a traditional FEM-based topology optimization code) or fake (design generated by the DL-based generator). The second discriminator is a regression-based deep neural network that counts the number of bars present in a design. The first one penalizes the quality and resolution of the designs generated by the generator and the second one penalizes the generator anytime it does not follow the geometrical condition (i.e. when the number of bars detected does not respect the inputted constraint). The generator is a ResUnet-based encoder decoder \cite{zhang2018road} that encodes the boundary and geometrical conditions and outputs the final 2D structure. 
We also train a CNN-based network to evaluate the mechanical performance of a generated design; given the design, the network outputs an approximation of the compliance value. The counter discriminator and compliance predictor are used to validate the generated designs, hence, contributing into the integration of  an automatic autonomous AI-based evaluation.


\section{Theoretical Overview/Background}\label{section:Background}

\subsection{Topology Optimization}\label{section:Topo}

Topology optimization seeks to find the optimal layout of a structure within a determined design space and for a specific set of load configurations and boundary conditions. It gained its success in the industrial world for its intrinsic characteristics: it allows an effective use of the material and has a higher degree of freedom when it comes to the design addressing the topology, shape and sizing problems all simultaneously. In the literature, we can find several approaches to solve a topology optimization problem: density-based\cite{bendsoe1989optimal}, level-set\cite{allaire2002level, allaire2004structural,wang2003level} and other methods. For further reading about the topology optimization approaches, one can refer to the review article of Sigmund and Maute \cite{sigmund2013topology} (2013). The top most common engineering commercial approach is the Solid Isotropic Material with Penalization (SIMP) method; also called the power-law approach\cite{sigmund1998numerical, zhao2010homogenization, zhou1991coc, mlejnek1992some}; introduced in 1989 by Bendsoe\cite{bendsoe1989optimal}. SIMP is a density gradient-based iterative method that uses penalization of the intermediate non-binary values of density material to converge to an optimal binary design. \\ SIMP approach is based on the following assumptions:
\begin{itemize}
  \item A design is represented by a specific distribution of discretized square material elements.
  \item Material properties are assumed constant within each element used to discretize the design domain. They are modelled as the relative material density raised to some power times the material properties of solid material.
  \item Variables are the element relative densities \(x_i\) . \(x_i\) represents either absence (0) or presence (1) of the material at each point of the design domain.
  \item A design is physically valid as long as the power \(p \geq 3\)\cite{bendsoe1999material} for Poisson's ratio \( = 1/3\) and is combined with a parameter constraint, a gradient constraint or filtering techniques.
\end{itemize}
A topology optimization problem based on the power law approach, where the objective is to minimize the compliance (i.e. the external work) \(c(x)\) can be written as the following:

\begin{equation}
min_x: = U^TKU = \sum_{e=1}^{N}x_e^p u_e^T k_0 u_e
\label{compliance_eq}
\end{equation}

subject to:

\begin{equation} 
\label{vol_frac_density_hook}
\begin{aligned}
\frac{V(x)}{V0}\leq f  \\ 
KU = F \\ 
0 < x_{min} \leq x \leq 1 
\end{aligned}
\end{equation}

where \(U\) and \(u_e\) are the global and element-wise displacements, \(F\) the forces vector, \(K\) and \(k_e\) are the global and element-wise stiffness matrices and N = number of elements used to discretize the design domain. \(x\) is the design variables vector i.e. the density material and \(x_{min}\) the minimum relative densities (non-zero to avoid singularity), \(p\) penalization power (typically 3). \(V_0\) and \(V(x)\) are the design domain volume and material volume respectively and \(f\) the volume fraction.
To efficiently solve the problem stated above, one approach is to adopt an iterative method that, for a specific design and its associated displacements, updates the design variables at each element of the FEM discretization independently from the updates at other elements. 
In the literature several approaches were proposed: the Optimality Criteria (OC) methods, Sequential Linear Programming (SLP)
methods or the Method of Moving Asymptotes (MMA). One simple approach is the OC method. Following OC, the design variables are updated as follows:

\begin{equation}
x_e = \left \{
    \begin{array}{lll}
        max(x_{min},x_e-m) & \mbox{if } x_e\beta^{\eta}_e \leq max(x_{min},x_e-m)\\
        x_e\beta^{\eta}_e & \mbox{if } max(x_{min},x_e-m) < x_e\beta^{\eta}_e < min(1,x_e+m)  \\
        min(1,x_e+m) & \mbox{if } min(1,x_e+m) \leq x_e\beta^{\eta}_e
    \end{array}
\right.
\label{OC_equation}
\end{equation}

where \(m\) is a positive   move limit, \(\eta=\frac{1}{2}\) is a numerical damping coefficient and \(\beta_e\) is found from the optimality condition as \(\beta_e = \frac{ \frac{-\partial c}{\partial x_e} }{\lambda \frac{\partial V}{\partial x_e} }\) such that $\lambda$ is a Lagrangian multiplier that can be found by a bi-sectioning algorithm and the sensitivity \(\frac{\partial c}{\partial x_e}\) of the objective function is found as \(\frac{\partial c}{\partial x_e} = -p x_e^{p-1} u_e^T k_0 u_e\).
To ensure the existence of solutions to the problem and avoid checker-board patterns\cite{sigmund1998numerical}, some restrictions should be introduced to the resulting design. One example is to apply a mesh-independency filter on the element sensitivities to ensure that the resulting design  is mesh-independent. Thus, the sensitivity of the objective function becomes \(\hat{ \frac{\partial c}{\partial x_e} } = \frac{1}{x_e \sum_{f=1}^{N}\hat{H_f}}\sum_{f=1}^{N} \hat{H_f} x_f \frac{\partial c}{\partial x_f} \) where the convolution operator (weight factor) \( \hat{H_f} =  \left \{
    \begin{array}{ll}
        r_{min}-dist(e,f) & \mbox{with } f \in N | dist(e,f) \leq r_{min} \mbox{ for } e= 1,...N \\
       0 & \mbox{outside the filter area. }
    \end{array}
\right. \) and \(dist(e,f)\) is the distance between the center of element \(e\) and the center of element \(f\).
In this study, the numerical solution of the SIMP method with the sensitivity analysis and optimality criterion written by Sigmund (2001)\cite{sigmund200199} is used to generate a sufficient database of 2D designs.
As we can see clearly, the geometric constraint controlling the complexity of the design cannot be easily incorporated in the topology optimization formulation. Hence, we make use of DL techniques and especially convolutional neural networks in order to integrate geometrical along with boundary conditions in the generation process. 

\subsection{Generative Adversarial Networks}\label{section:gan}

Generative adversarial networks GAN were first introduced by Ian Goodfellow in his paper\cite{goodfellow2014generative}. This method learns to mimic any input data distribution. One advantage of GANs is that it helps increase the amount of data by generating new samples that follow the same distribution of the original existing ones.
A GAN consists of two neural networks the generator and the discriminator, each working against the other. The generator tends to output samples with a distribution similar to the original ones. However, the discriminator tries to discriminate real sample (i.e. coming from the original data) from synthesized ones (i.e. coming from the generator). The main purpose of a GAN is to optimize the generator to produce creative data samples of large variety without memorizing the original samples. 
Mathematically speaking, the generator wants to learn a density distribution \(p_g\) that is similar to the real data distribution \(p_{data}\) by updating its mapping function \(G(z, \theta_g)\); such that \(\theta_g\) are the parameters of the generator function and \(z\) is a latent random vector following a noise prior distribution \(p_z\). The discriminator \(D(x, \theta_d)\) wants to output the probability that \(x\) came from the real data rather than \(p_g\); such that \(\theta_d\) are the parameters of the discriminator function. Both networks are trained in a minimax framework to improve the same loss function: the binary log loss \(L(G,D)\). This loss, as defined in the original paper\cite{goodfellow2014generative} is:

\begin{eqnarray}
 L(G,D) &=& \underset{G}{\min}~ \underset{D}{\max}~ \mathbb{E}_{x \sim p_{\text{data}(x)}} [log(D(x))] \notag\\
&+& \mathbb{E}_{z\sim p_z(z)} [log(1-D(G(z))) ]
\label{binary_log_loss_eq}
\end{eqnarray}

The solution of this function is \(p_g = p_{data}\) i.e. when the generator starts to output data samples following the same distributions as the real samples. \\
On the other hand, a conditional GAN (cGAN) \cite{mirza2014conditional} is an extension of the GAN network enabling the generation to be oriented by a specific input condition \(c\). In this framework, the basics of cGAN become: the conditional generator as \(G((z/c),\theta_g)\), the conditional discriminator as \(D((x/c),\theta_d)\) and the loss function as: 

\begin{eqnarray}
L(G,D) &=& \underset{G}{\min}~ \underset{D}{\max}~ \mathbb{E}_{x \sim p_{\text{data}(x)}} [log(D(x/c))] \notag\\ &+& \mathbb{E}_{z\sim p_z(z)} [log(1-D(G(z/c))) ]
\label{conditional_binary_log_loss_eq}
\end{eqnarray}
The approach adopted in this work is based on the cGAN framework. 

\section{Methodology}\label{section:method}
\begin{figure*}
\begin{center}
\includegraphics[width=16cm]{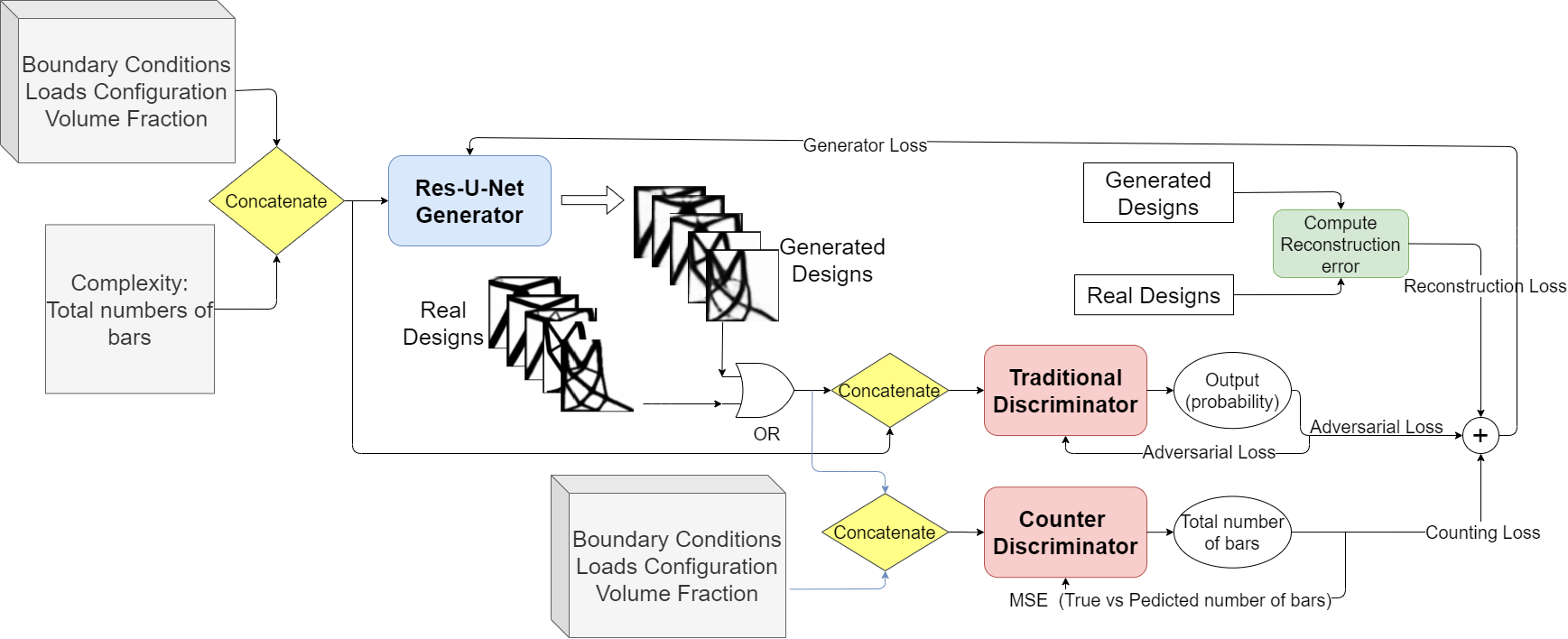}
\caption{TRAINING PROCEDURE. \label{figure_Training_Process_} }
\end{center}
\end{figure*}
\begin{figure}
\begin{center}
\includegraphics[width=6cm]{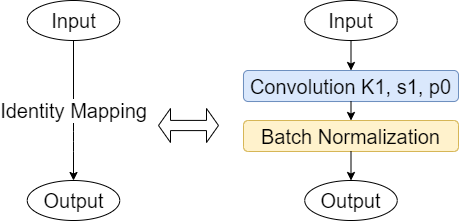}
\caption{IDENTITY MAPPING. An identity mapping connection consists of a convolution of kernel size\(=1\times1\), stride of 1 and padding of 0 followed by a batch normalization layer.
\label{figure_Identity_Mapping} } 
\end{center}
\end{figure}
\begin{figure}
\begin{center}
\includegraphics[width=6cm]{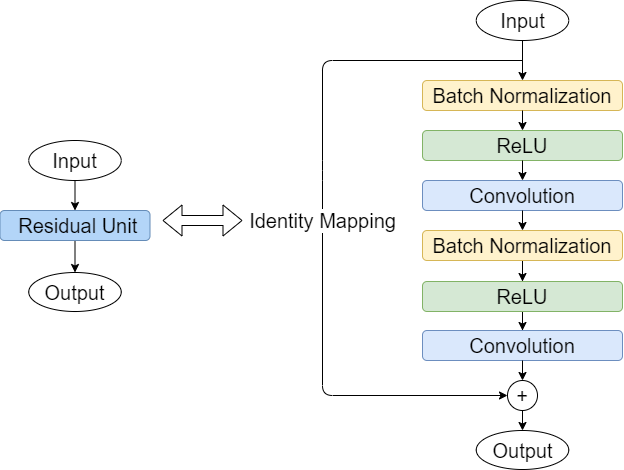}
\caption{RESIDUAL UNIT.,
\label{figure_Residual_unit} }
\end{center}
\end{figure}
\begin{figure}
\begin{center}
\includegraphics[width=9cm]{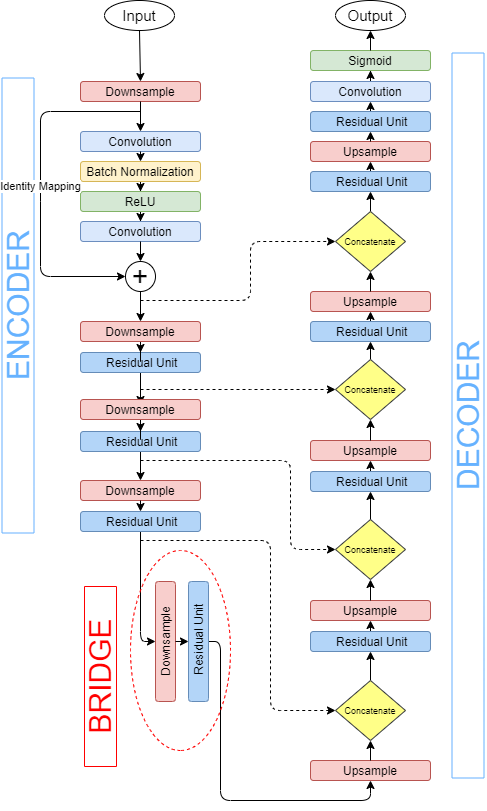}
\caption{RES-U-NET GENERATOR'S ARCHITECTURE.,
\label{figure_GEN} }
\end{center}

\end{figure}
\begin{figure}
\begin{center}
\includegraphics[width=6cm, height=13cm]{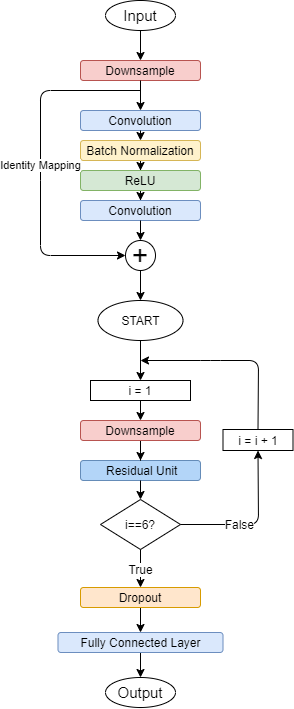}
\caption{RESIDUAL TRADITIONAL DISCRIMINATOR'S ARCHITECTURE.,
\label{figure_traditional_DISC} }
\end{center}
\end{figure}
\begin{figure}
\includegraphics[width=10cm, height=8cm]{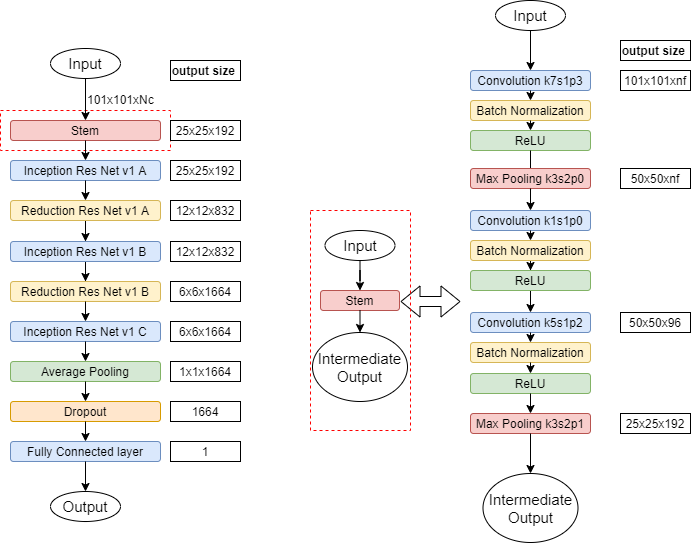}
\caption{INCEPTION RES-NET V1 COUNTER DISCRIMINATOR'S ARCHITECTURE. The number of input channels \(Nc\) = 6; the input to the counter discriminator is a six-channel image: the design, the boundary conditions (\(BC_x\) and \(BC_y\)), the loads configurations (\(F_x\) and \(F_y\)) and the volume fraction. The number of feature maps \(nf\) = 32.
\label{figure_COUNT_DISC} }
\end{figure}

In this work, a conditional \cite{mirza2014conditional} convolutional\cite{radford2015unsupervised} dual-discriminator GAN has been adopted, especially since the input conditions are shaped as images (Fig.~\ref{figure_input_generator}): boundary and loads conditions, volume fraction and the complexity i.e. number of bars in a design and the outputs are 2D structures that can be represented as 2D images. This approach consists of a deep ResUnet generator\cite{zhang2018road}, a CNN residual-based discriminator to differentiate between the real designs and the generated ones and an inception-based\cite{szegedy2017inception} bar counter to quantify the complexity of the generated/real designs.
The training procedure is detailed in Fig.~\ref{figure_Training_Process_}. The code is available on \href{https://github.com/waadALMASRI/SHAPE-RELATED-CONSTRAINTS-AWARE-GENERATION-OF-MECHANICALDESIGNS-THROUGH-DEEP-CONVOLUTIONAL-GAN.git}{this GIT repository}\footnote{The GIT repository will become pubic when the paper will be accepted.}.
\subsection{Architecture of the generator}

The generator is a deep ResUnet\cite{zhang2018road} network. It’s an encoder-decoder convolutional architecture with residual and skip connections between the outputs of the encoder layers and the inputs of the decoder layers or what is called U-Net. This architecture benefits from both: the U-Net and residual advantages. In fact, the U-Net\cite{ronneberger2015u} improves the information propagation from the encoder to the decoder and compensates to the loss of finer details in the decoding process by combining low level features to their corresponding high levels. Although U-Net facilitates the flow of information, the deeper the network gets, the worse it performs. This degradation in performance with deep neural networks, usually due to vanishing gradients, is overcome with the deep residual learning framework proposed by He et al. (2016)\cite{he2016deep}. Instead of learning a certain mapping H(x), we learn its residual alternative i.e. F(x)+x such that x is the identity mapping of the input x (Fig.~\ref{figure_Identity_Mapping}, Fig.~\ref{figure_Residual_unit}). In their paper, He et al.\cite{he2016deep} justifies this choice by stating that it is easier to optimize the residual mapping than to optimize the original one especially if the identity mapping was optimal. Hence, it would always be easier for the network to push the residual to zero than to fit an identity mapping by a stack of nonlinear layers. In addition, This ResUnet architecture was also used in previous work\cite{kallioras2020accelerated} and achieved promising performance on generating nonlinear fine-detailed 2D structures from input boundary conditions.
The network can be divided to three parts: an encoder, a bridge and a decoder. The encoder is formed of 4 blocks each consisting of a down-sampling layer (a convolution of stride 2) and a residual unit as shown in Fig.~\ref{figure_Residual_unit}. The decoder is formed of 5 blocks each consisting of an up-sample layer (a transpose convolution of stride 2) and a residual unit followed by a convolution of kernel size \(1\times1\) and a sigmoid activation. The bridge connection has the same architecture as an encoder’s block and combines the encoder to the decoder. The diagram of the full architecture is shown in Fig.~\ref{figure_GEN} and the information about each layer's kernel size, output size, etc. are detailed in the table Tab~\ref{table_GEN_Architecture} in the ~\ref{appendix:A}. 

\subsection{Architecture of the discriminators}
The first one i.e. the traditional discriminator takes as input the design along with the geometrical and mechanical conditions and outputs the probability that the design comes from the real data distribution to ensure a better reconstruction of the design.  The second one is a regression-based DL counter. Its architecture was inspired from the Inception ResNet v1 modules\cite{szegedy2017inception}; they allow in increasing the network complexity (more units per layer) without generating any extra computational costs. It takes as input the design and only its corresponding mechanical conditions and outputs the number of total bars present in the design to ensure that the generated design respects the input geometrical condition.

\subsubsection{Architecture of the traditional discriminator}
The network of the traditional discriminator follows the generator's encoder architecture and is detailed in Fig.~\ref{figure_traditional_DISC}.

\subsubsection{Architecture of the counter discriminator}
The counter network detailed in Fig.~\ref{figure_COUNT_DISC} consists of a stem, an Inception Resnet v1 block A, a Reduction Resnet v1 block A, an Inception  Resnet v1 block B, a Reduction Resnet v1 block B, an Inception Resnet v1 block C followed by an average pooling layer, a dropout layer and a fully connected layer. The inception/reduction blocks used defers from the original paper\cite{szegedy2017inception} by the number of input/output feature maps.
In this work, the counter is pre-trained to predict the total number of bars present in the real 2D designs (SIMP-based designs) before the full training of the GAN, for this procedure makes the accuracy in predicting the number of bars in the reconstructed designs more precise, therefore, improves the training of the generator. 

\subsection{Loss Function}

The objective of this dual-discriminator GAN is to train a generator that covers two aspects: the quality of the reconstructed 2D structures and their conformity to the mechanical and geometrical conditions. Thus, the original adversarial loss function (Eq.~\ref{conditional_binary_log_loss_eq}) used to train the generator was altered to consider both aspects: a reconstruction loss and a counting loss were added to the formulation. The generator loss function adapted in the training is the following:

\begin{equation}
L_G = \lambda_1 L_{reconstruction} + \lambda_2 L_{adversarial} +  \lambda_3 Acc_{counter} L_{counting}
\label{eq_gen_loss}
\end{equation}

such that:

\begin{alignat}{1}
L_{adversarial} = E_{z\sim p_z(z)} [log(1-D(G(z/c))) ] \label{eq_gen_adversarial_loss}\\
L_{reconstruction} = \frac{1}{N} \sum_{i=1}^{N}(x_i - \hat{x}_i)^2 \label{eq_gen_recnstruction_loss}\\
L_{counting} = \frac{1}{N} \sum_{i=1}^{N}(\hat{y}_i \geq y_i) 
\label{eq_gen_counting_loss}
\end{alignat}

with \(x_i\) and \(\hat{x}_i\) the true and predicted 2D structure respectively, the \(y_i\) and \(\hat{y}_i\) the true (i.e. the input geometrical condition) and predicted (i.e. the number of bars in the generated designs) maximum number of total bars respectively, \(N\) the batch size and \(Acc_{counter}\) the accuracy of the discriminator counter on the real samples\footnote{\(Acc_{counter} = \frac{1}{N} \sum_{i=1}^{N}(\hat{t}_i = y_i) \) with \(\hat{t}_i\) the predicted number of bars present in the real designs}. 
The adversarial loss (Eq.~\ref{eq_gen_adversarial_loss}) ensures creative and varied 2D structures. The reconstruction loss (Eq.~\ref{eq_gen_recnstruction_loss}) boosts the aesthetics and the reproduction of high-frequency details in the generated samples.
The counting loss (Eq.~\ref{eq_gen_counting_loss}) penalizes the generator every time the upper bound constraint on the total number of bars is not respected\footnote{The expression \( (\hat{y}_i \geq y_i) \) is equivalent to  the count of reconstructed designs having a total number of bars \(\hat{y}_i \) greater than the input condition \(y_i\)}.
In this work,\( \lambda_1 \), \( \lambda_2 \) and \( \lambda_3 \) are set to 1, 0.01 and 0.1 respectively in order to have the same order of magnitude.

\subsection{Compliance Predictor}\label{section:compliance}

\begin{figure}
\begin{center}
\includegraphics[width=10cm]{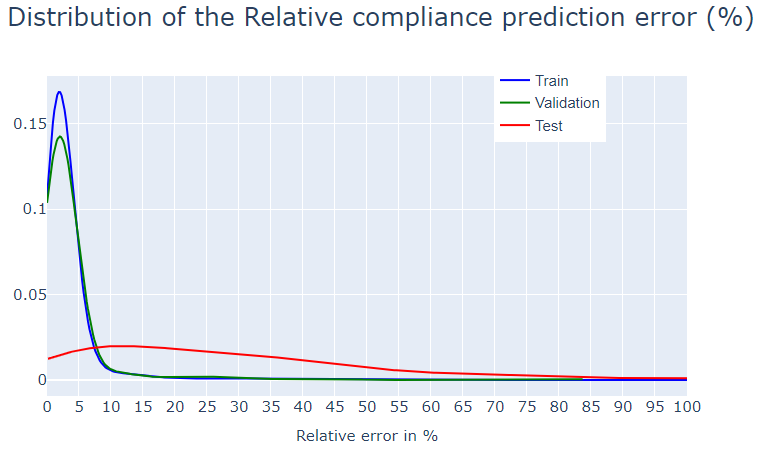}
\caption{DISTRIBUTION OF THE RELATIVE COMPLIANCE PREDICTION ERROR.
\label{figure_compliance_prediction_accuracy} }
\end{center}

\end{figure}

\begin{figure}
\begin{center}
\includegraphics[width=10cm]{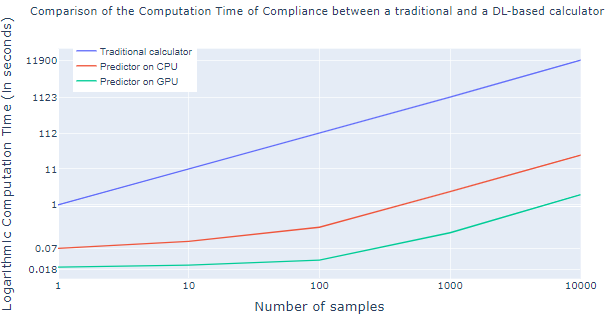}
\caption{COMPARISON OF COMPUTATION TIME (IN SECONDS) BETWEEN TRADITIONAL COMPLIANCE CALCULATOR AND DL-BASED COMPLIANCE PREDICTOR.
\label{figure_compliance_time_prediction} }
\end{center}
\end{figure}

In order to automate and accelerate the full topology optimization process, a DL-based mechanical validator was also built: the compliance predictor. It is a deep residual CNN-based network which takes the design as input and outputs an approximation of the mean compliance value. The compliance predictor is a Resnet-SE-based\cite{Hu_2018_CVPR} deep convolutional network. 
To evaluate this compliance predictor, we consider two aspects: the error and the computation time of the prediction. As reported in Fig.~\ref{figure_compliance_prediction_accuracy} and Tab.~\ref{table_compliance_accuracy},   93\% of the predictions over the validation set are made within 10\% of error margin and 87\% of the predictions fall into a 5\% error margin. The error calculated is the relative percentage error between the true and predicted values: \( e_{\%} = \frac{|True-Predicted|}{True}\times100\). A compliance FE-based computation (coded in pytorch) takes 1.13 seconds per design to output the result. However, a pytorch Resnet compliance predictor needs 0.07 seconds per design on CPU and 0.021 seconds on one GPU to compute an approximation of the compliance, i.e. 16 times (on CPU) and 56 times (on GPU) faster per design and 400 (on CPU) to 5000 (on GPU) times faster for a larger number of designs. Thus, one can evaluate 1 or 10 000 designs in simply few seconds. The table (Tab.~\ref{table_compliance_time_prediction}) and figure (Fig.~\ref{figure_compliance_time_prediction}) shown below compare the computation time needed for every type of predictor to compute the compliance values of \(n\) samples. \\
One abnormal behavior of the compliance predictor was detected when predicting the compliance over the test set (i.e. the inner loads-based designs, refer to the Section~\ref{section:data_generation}): the accuracy of the predictor was unacceptable; only 25\% of predictions fell under a margin error of 10\%. After checking the distribution of the compliance values in the different sets (Fig~\ref{figure_compliance_values_distributions_in_sets}), we have identified that the test set showed a mean and a standard deviation of the compliance values twice greater than the ones seen in the train and validation sets. Further investigations shall be done in order to compensate for this pitfall. \\
Consequently, in this study, we will be using the compliance predictor to validate only the edge-loads based designs while we will compute the compliance using the FE-method for the inner-loads-based designs i.e. the test set designs. \\
In the future, the compliance predictor module can be added as a third discriminator in order to also penalize the generator over the compliance. Moreover, this CNN-based compliance predictor could also replace its calculator version in the SIMP code in order to accelerate the process.
\begin{table}[t]
\caption{COMPLIANCE PREDICTION ACCURACY. This table displays the accuracy of the Resnet-SE-based network in predicting the compliance value based on the design as input. The accuracy is defined as the percentage of designs having the error margin below \(2.5\%\), \(5\%\), \(7.5\%\) and \(10\% \). The error margin is defined as the following:
 \( e_{\%} = \frac{|C_i-C_g|}{C_i}\times100\) such that \(C_g\) is the predicted compliance values and \(C_i\) is the true compliance values (the ground truth). }
\label{table_compliance_accuracy}
\begin{center}
{\small
\begin{tabular}{ |p{1.5cm}|p{2cm}|p{2cm}|p{2cm}|p{2cm}|}
\hline Dataset &  \(e_{\%} \leq 2.5\%\) &  \(e_{\%} \leq 5\%\)&  \(e_{\%} \leq 7.5\%\) & \(e_{\%} \leq 10\%\)\\
\hline Train & 54.28\% & 86.28\% & 92.71\% & 94.23\%  \\ 
\hline Validation & 53.47\% & 87\% & 92.6\% & 93.75\%  \\
\hline Test & 7\% & 13\% & 19.5\% & 25.35\% \\
\hline
\end{tabular}
}
\end{center}
\end{table}
\begin{table}[t]
\caption{COMPARISON OF COMPUTATION TIME (IN SECONDS) BETWEEN TRADITIONAL COMPLIANCE CALCULATOR AND DL-BASED COMPLIANCE PREDICTOR.}
\label{table_compliance_time_prediction}
\begin{center}
{\small
\begin{tabular}{ |p{2cm}|p{1cm}|p{1cm}|p{1cm}|p{1cm}|p{1cm}|}
\hline Number of samples \(n\) & 1 & 10 & 100 & 1000 & 10 000 \\
\hline Traditional calculator & 1.13 & 11.24 & 112  & 1123.6 & 11 900 \\
\hline DL-based predictor on CPU & 0.07 & 0.11 & 0.27 & 2.66 & 27.21 \\
\hline DL-based predictor on 1 GPU & 0.021 & 0.024 & 0.033 & 0.19 & 2.19 \\
\hline
\end{tabular}
}
\end{center}
\end{table}


\section{Experiments and Results}\label{section:results}

\subsection{Data Generation}\label{section:data_generation}

\begin{figure}
\includegraphics[width=16cm]{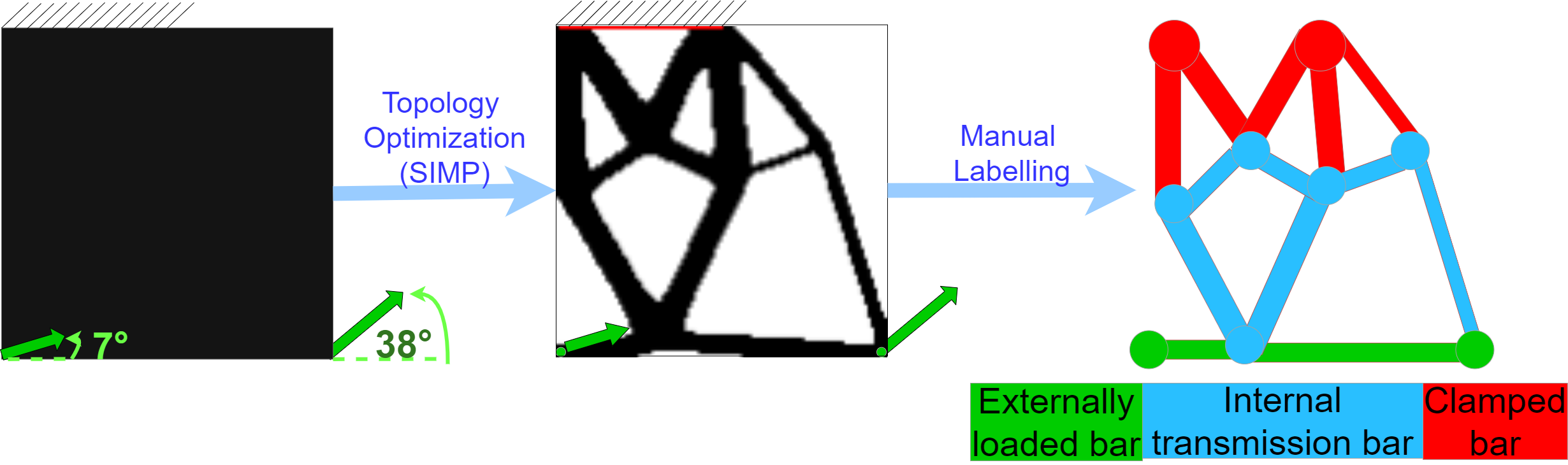}
\caption{TYPES OF BARS IN A DESIGN. The design shown in this figure is clamped from upper left edge and loaded with 2 external concentrated forces located in the bottom corners. The bottom left load’s orientation is \(7^\circ\) and the bottom right one is of \(38^\circ\). As we can see, the design has 5 clamped (red) bars outgoing from two nodes, 2 externally loaded (green) bars outgoing from two nodes corresponding to the location of the forces and 6 internal transmission (bleu) bars, thus, a total of 13 bars.
\label{figure_Bar_Labelled_design} }
\end{figure}

\begin{figure}
\includegraphics[width=16cm]{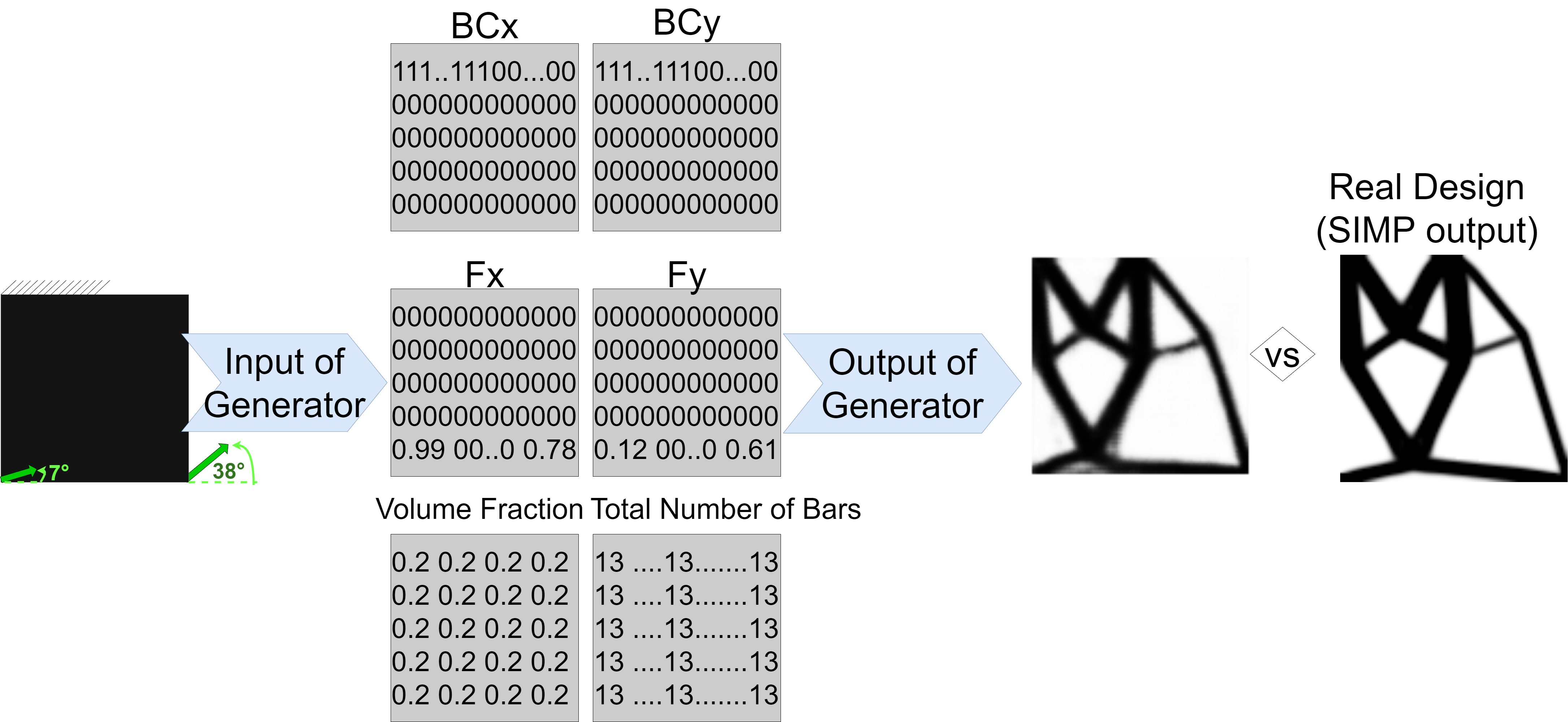}
\caption{EXAMPLE OF INPUT TO GENERATOR. Boundary conditions, loads configuration, volume fraction and complexity (i.e. total number of bars) are formulated as a six-channel image. \(BC_x\) and \(BC_y\) are the fixed nodes matrices. \(F_x\) and \(F_y\) are the loaded nodes matrices.
\label{figure_input_generator} }
\end{figure}

To train the model, 2D designs were generated following the SIMP method explained in section~\ref{section:Topo} via an academic open-source topology optimization code written by Sigmund\cite{sigmund200199} by varying the input BCs, load configurations, volume fraction, etc.
A 2D structure of size \(n_x\times n_y\) can be discretized into a mesh of \((n_x+1)\times(n_y+1)\) nodes and is subject to 2 major constraints: the boundary conditions i.e. the fixed nodes and the load configurations i.e. the loaded nodes. In this work, every SIMP-based design is a \(100\times100\) pixels black and white image and is obtained corresponding to a random set of configurations. Both the design and its configurations were saved for the training and testing stages of the GAN. To generate a wide variety of designs, we have sampled the boundary and load configuration using the following strategy:
\begin{itemize}
  \item The volume fraction follows a normal distribution of mean of 0.3 and standard deviation of 0.05: \(V \sim N(0.3, 0.05)\)
  \item The number of loads follows a Poisson distribution of \(\lambda=2\): \(N_L \sim P(\lambda=2)\)
  \item The orientation of the loads follows a uniform distribution between 0 and 360 degrees: \(\theta_i \sim U[0, 360^\circ]\) with \(i = 1, ... N_L\)
  \item The number of fixed nodes follows a Poisson distribution of \(\lambda=50\): \(N_{BC} \sim P(\lambda=50)\)
  \item In this study we limit the locations of fixed and load nodes to the edge ones such that fixed and loaded nodes are always of opposite sides except for the test set. In the latter, internal (non-edge) loads were considered.
\end{itemize}
A set of boundary conditions are represented as \((n_x+1)\times(n_y+1)\) matrices with null values everywhere except for the fixed nodes set to 1.0; for simplicity, we only consider encastrated designs i.e. the boundary conditions along the \(x\) axis (\(BC_x\)) and \(y\) axis (\(BC_y\)) are similar. 
Loads configurations \(F_x\) and \(F_y\) are represented as \((n_x+1)\times(n_y+1)\) matrices with null values everywhere except for the loaded nodes; a loaded node \(n_e\) located at line \(i\) and column \(j\) tilted \(\theta\) degrees has \(F_x(i,j) = cos(\theta)\) and \(F_y(i,j) = sin(\theta)\); the magnitude of the loads were set to \(1.0\) \(N\).
Not all designs outputted by SIMP were used in this study, only truss-like structures were kept to train the GAN-based approach i.e. designs with bar-like structures. 
As explained previously, the traditional discriminator takes as input the design along with the boundary, load, volume fraction and complexity configurations. Thus, in order to concatenate all these inputs together: the 2D design is reshaped into a \(101\times101\) image, the volume fraction and the complexity are represented as \(101\times101\) matrices of the same value equal to the volume fraction and the number of bars respectively; an example of an input to the generator is shown in Fig.~\ref{figure_input_generator}. Consequently, the traditional discriminator input and the counter input are seven-channel and six-channel images respectively.
The additional information i.e. the complexity was added by a manual labeling. Every design is fixed from an edge and loaded from another; hence, any 2D structure can be represented by the combination of three types of bars: the clamped bars i.e. fixed, the externally loaded bars and the internal transmission bars i.e. neither clamped nor loaded. The figure~\ref{figure_Bar_Labelled_design} shows an example of a design annotated, with a red line highlighting its fixed edge and green dots showing the location of the forces, along with its bar-type-labeled counterpart.
The dataset was separated into train (3885 samples), validation (432 samples) and test sets (635 samples). The only difference between the train/validation sets and the test set is that in the latter, internal loads were considered in order to test the generalization aspect of the generator.\\
{\it NB: in this work, the validation set has the same data distribution as the training set (i.e. all validation designs have only edge loads). However, this validation set does not take part in the training procedure. It is used to test the generator's performance over the designs having the same distribution as the trained ones.}

\subsection{Results}\label{section:results_}

It is rare to find any objective evaluation for GANs in the literature, especially that its evaluation correlates with the type of data generated. Thus, most of the evaluations are subjective and based on aesthetics of the generated samples. In this study, we don't only test the performance of the generator on creativity and aesthetics but also on complying with the wanted specifications: the complexity and the mechanical validity. The mechanical validity is defined by the compliance value and the volume fraction constraint while the complexity is defined by the total number of bars present in the generated structure. These evaluation metrics are solid material for decision making when it comes to finding a compromise between mechanical performance and manufacturability, represented here by the complexity of the design. Thus, in the following sections, we will detail the evaluation methodology and present the results.

\subsubsection{Counter Discriminator's performance}\label{section:counter_results}

\begin{table}[t]
\caption{COUNTING ACCURACY. This table displays the accuracy of the counter discriminator in predicting the total number of bars in a design. The Accuracy (\(Acc\)) stands for the percentage of predictions matching perfectly the true labels. The Accuracy$\pm$1 bar (\(Acc_{\pm 1\hspace{1mm}bar}\)) stands for the percentage of predictions with an error of 1 bar more or less than the true labels. The  Accuracy$\pm$2 bars (\(Acc_{\pm 2\hspace{1mm}bars}\)) stands for the percentage of predictions with an error of 2 bars more or less than the true labels. }
\label{table_counter_accuracy}
\begin{center}
{\small
\begin{tabular}{ |p{1.2cm}|p{1.2cm}|p{1.2cm}|p{1.2cm}|}
\hline Dataset &  \(Acc\) &  \(Acc_{\pm 1\hspace{1mm}bar}\) & \(Acc_{\pm 2\hspace{1mm}bars}\) \\
\hline Train & 92.89\% & 99.87\% & 100\%  \\
\hline Validation & 55.55\% & 85.41\% & 94.9\% \\
\hline Test & 26.93\% & 67.71\% & 87.24\% \\
\hline
\end{tabular}
}
\end{center}
\end{table}
In order to test the conformity of the generator to the complexity constraint, the counter discriminator trained in our GAN-based network is used to predict the number of bars in the generated designs. The Accuracy of this discriminator is assessed using the original designs (designs generated by the SIMP approach) since we already know their true complexities (i.e. total number of bars) from the manual labeling.
Table~\ref{table_counter_accuracy} presents the accuracy of the counter in predicting the exact number of bars in a design \(Acc\), the accuracy of predicting the bars with more or less 1 bar counted \(Acc_{\pm1bar}\) and the accuracy of predicting the bars with more or less 2 bars counted \(Acc_{\pm2bars}\) for the train, validation and test sets. From this table, we spotted one major observation: the large gap between the \(Acc\) and the \(Acc_{\pm1bar}\) in the validation and test sets. This comes from the fact that the range of values of the internal transmission bars is very wide with respect to the other types (the clamped and the loaded bars) and not well represented in the dataset. Figure~\ref{figure_type_of_bars_distributions} (in the ~\ref{appendix:A}) plots the distribution of every type of bars. It shows that a large proportion of designs lack of internal transmission bars and when this type of bars is present, it ranges from 1 to 20 bars with a high proportion between 1 and 5 bars, which makes it hard to predict its value with precision, therefore, justifying the gap between \(Acc\) and \(Acc_{\pm1bar}\). However, in our case, predicting the complexity with more or less 2 bars is admissible. Moreover, Table~\ref{table_counter_accuracy} shows great performance on the validation set (the accuracy of predicting the complexity with more or less 2 bars (\(Acc_{\pm2bars}\)) is 94.9\%) and on the test set ( \(Acc_{\pm2bars}\) is 87.24\%).
Consequently, in the next section~\ref{section:generator_results}, the complexity constraint will be evaluated on the generated designs considering the DL inception-based counter discriminator. \\
{\it
We note that the generated designs are not transformed in any way (i.e. no threshold, nor erosion/dilation was applied on the generated designs) prior to the prediction of the complexity. }

\subsubsection{Generator's performance}\label{section:generator_results}

\begin{figure*}

\includegraphics[width=16cm]{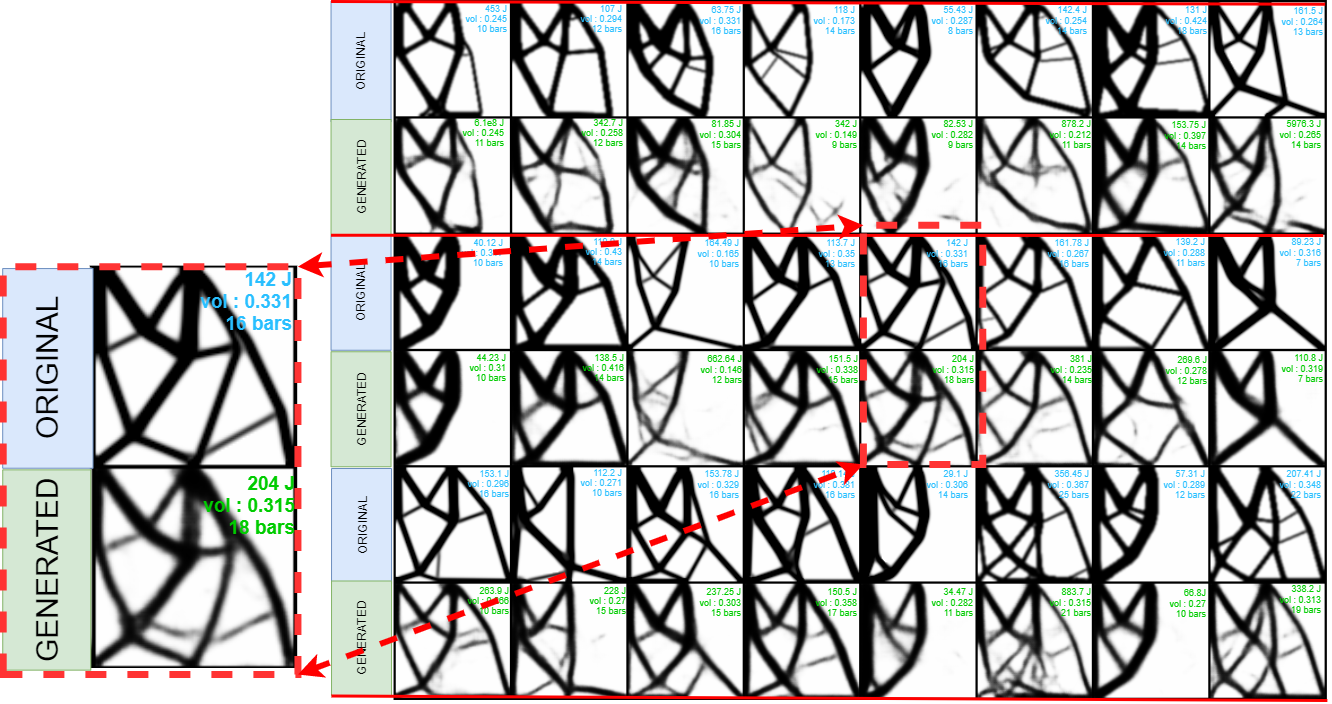}
\caption{VALIDATION SET RESULTS. All designs present in this figure are fixed from the left half of the upper edge and loaded in the opposite bottom edge with different number of loads and orientations. We compare for a set of BCs, loads configuration, volume fraction and complexity: the shape, the compliance (in Joules \(J\)) and the constraints of complexity \((bars)\) and volume fraction \((vol)\) of the original design versus the generated one.
\label{figure_validation_set_results_1} }
\end{figure*}
\begin{figure}
\begin{center}
\includegraphics[width=16cm]{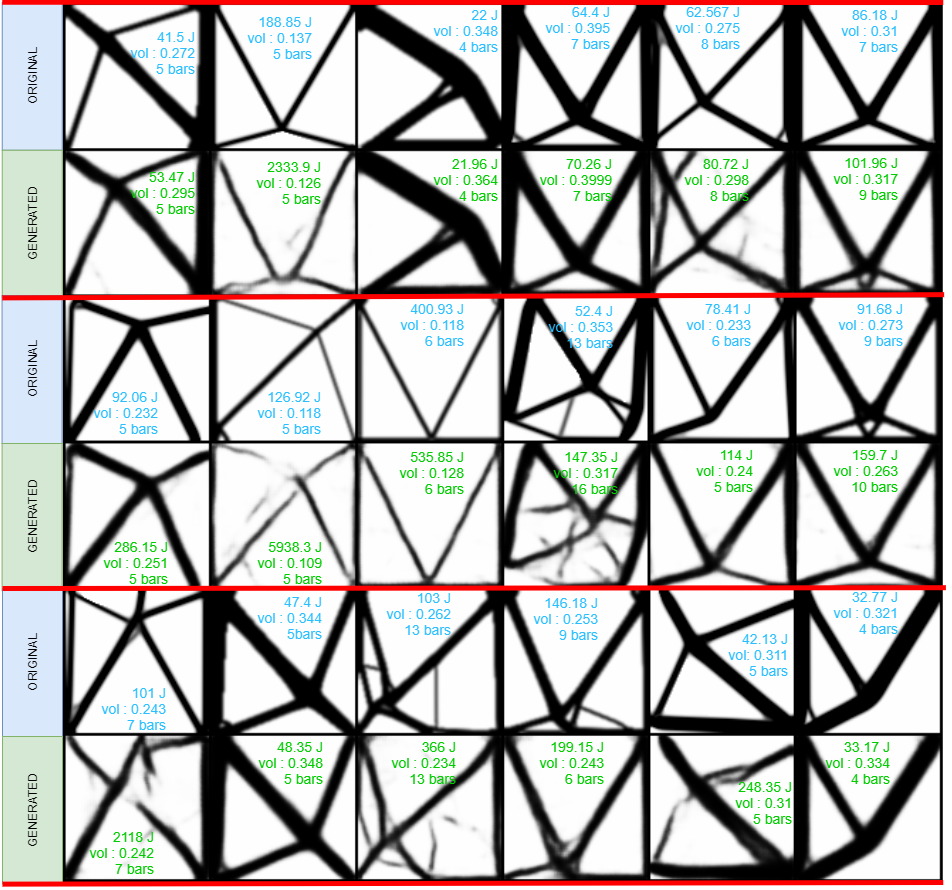}    
\end{center}
\caption{VALIDATION SET RESULTS - 2. All designs present in this figure are either fixed from the the upper edge and loaded in the opposite bottom edge or inversely with different number of loads and orientations.
\label{figure_validation_set_results_2} }
\end{figure}
\begin{figure}
\includegraphics[width=16cm]{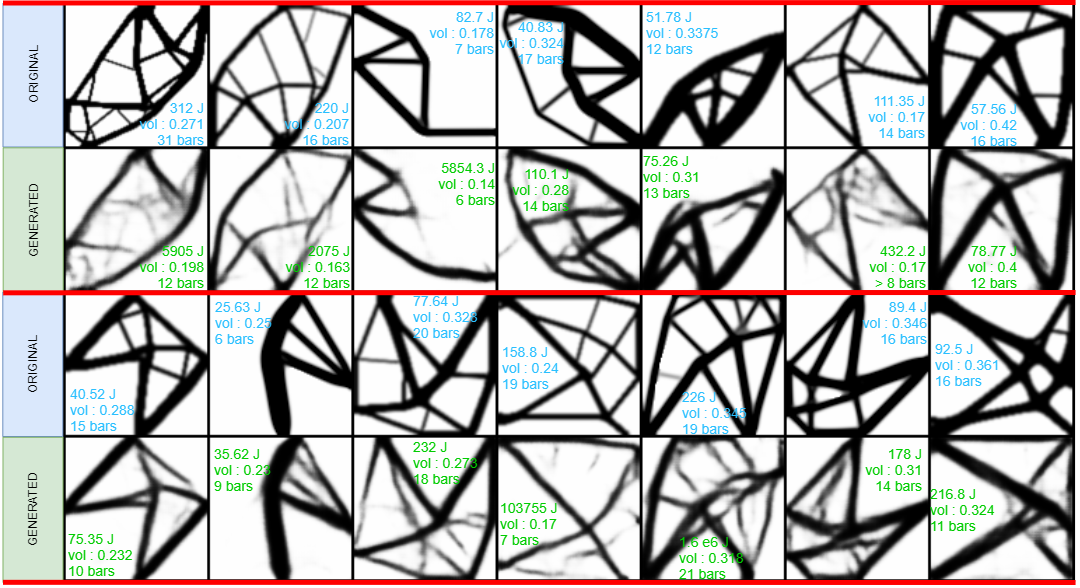}
\caption{COMPLEX DESIGNS - VALIDATION SET RESULTS.
\label{figure_validation_set_results_3} }
\end{figure}
\begin{figure*}
        \includegraphics[width=16cm]{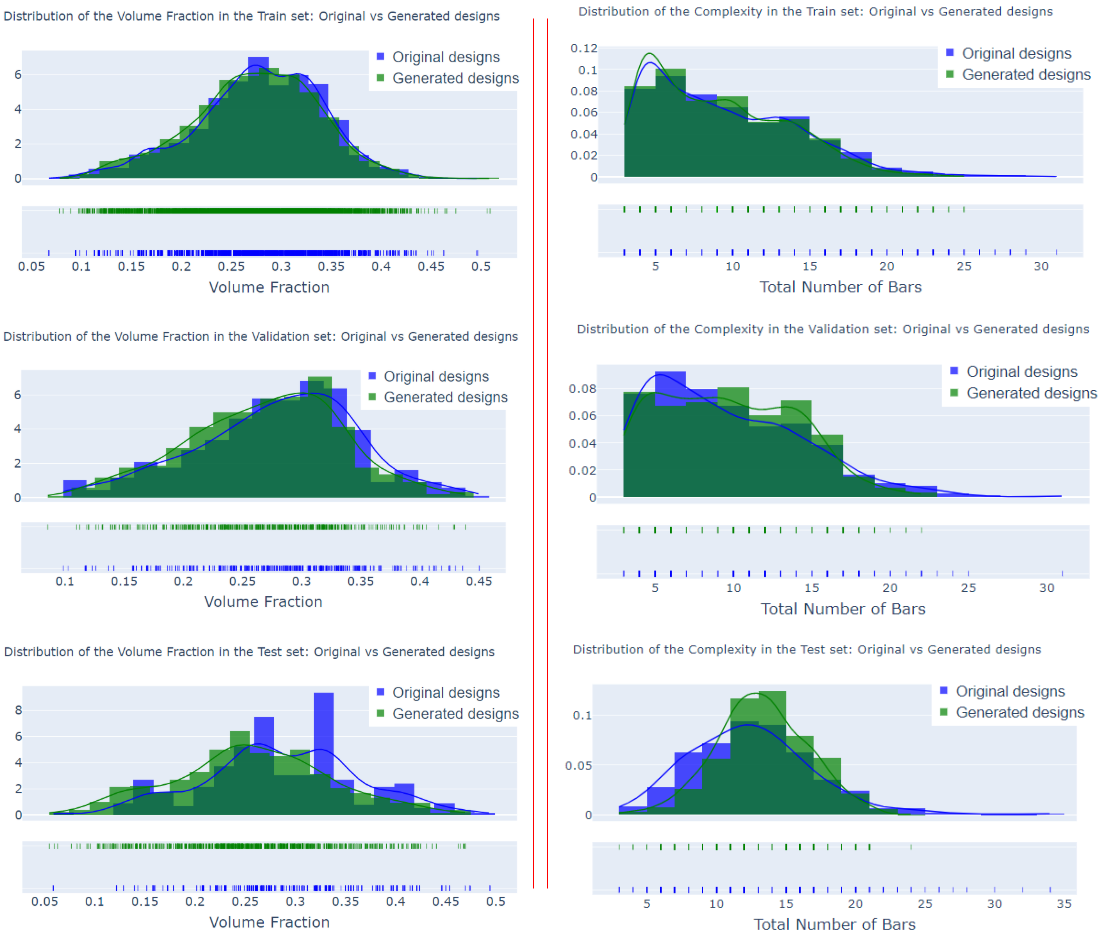}
\caption{COMPLEXITY AND VOLUME FRACTION DISTRIBUTIONS IN THE TRAIN, VALIDATION AND TEST SETS. On the left side, the distributions of the volume fractions of the generated versus original (SIMP-based) designs are shown. On the right side, the distributions of the complexity of the generated versus the original designs.
\label{figure_distributions}}

\end{figure*}
\begin{figure*}
        \includegraphics[width=16cm]{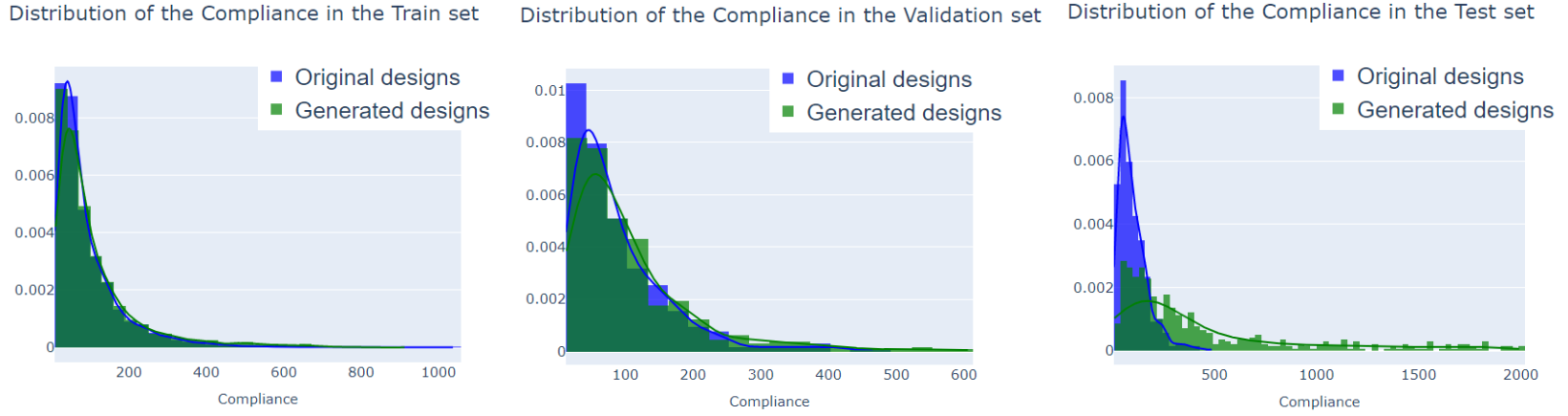}
\caption{DISTRIBUTIONS OF THE COMPLIANCE OF THE GENERATED VS SIMP-BASED DESIGNS IN THE TRAIN, VALIDATION AND TEST SETS. In this figure, the distributions of the compliance values of the generated versus original (SIMP-based) designs are shown. NB: the compliance values of the generated designs of the train and validation sets were computed using the CNN-based compliance predictor while the compliance values of the test set were computed using the FE-based compliance calculator {\it For the sake of the visualization, compliance values higher than \(2000 J\) were omitted.}.
\label{figure_compliance_distributions} }
\end{figure*}
\begin{figure}
\includegraphics[width=16cm]{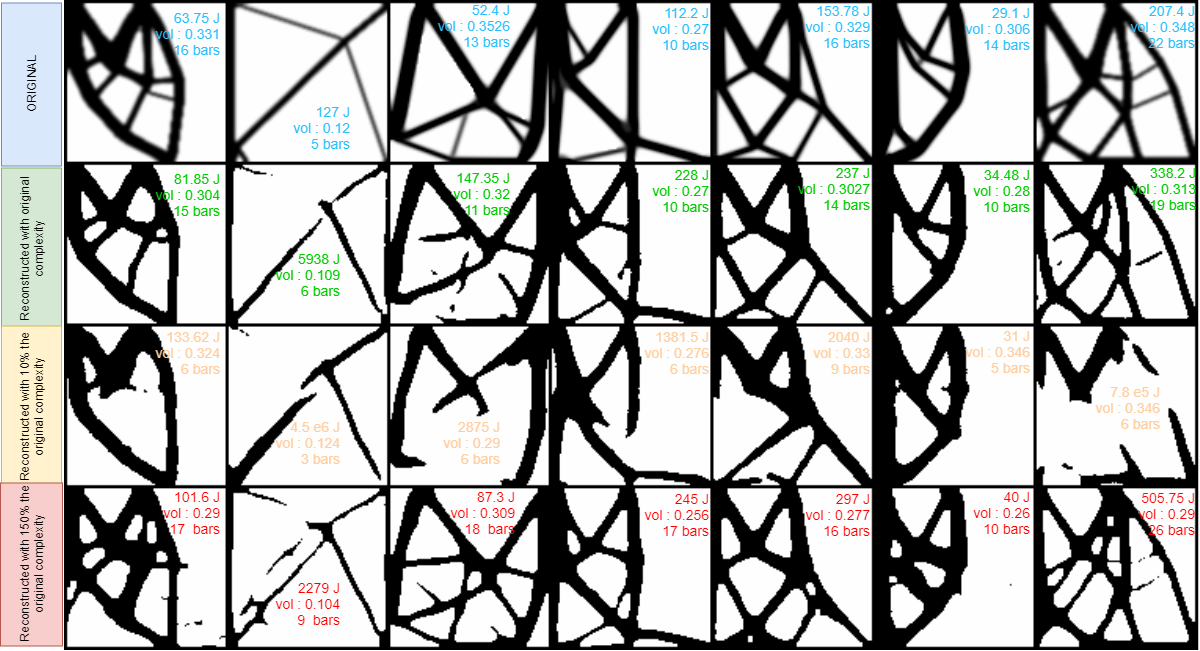}
\caption{DESIGNS WITH CHANGING COMPLEXITIES. In this figure, seven designs were generated with the original complexity (the total number of bars of the original design \(N\)) and with changing complexities: 10\% of the complexity (\(0.1\times N\) bars) and 150\% of the complexity (\(1.5\times N\)) bars; for a better visualization, we have threshold the designs.
\label{figure_complexity_variation} }
\end{figure}
\begin{figure}
        \includegraphics[width=16cm]{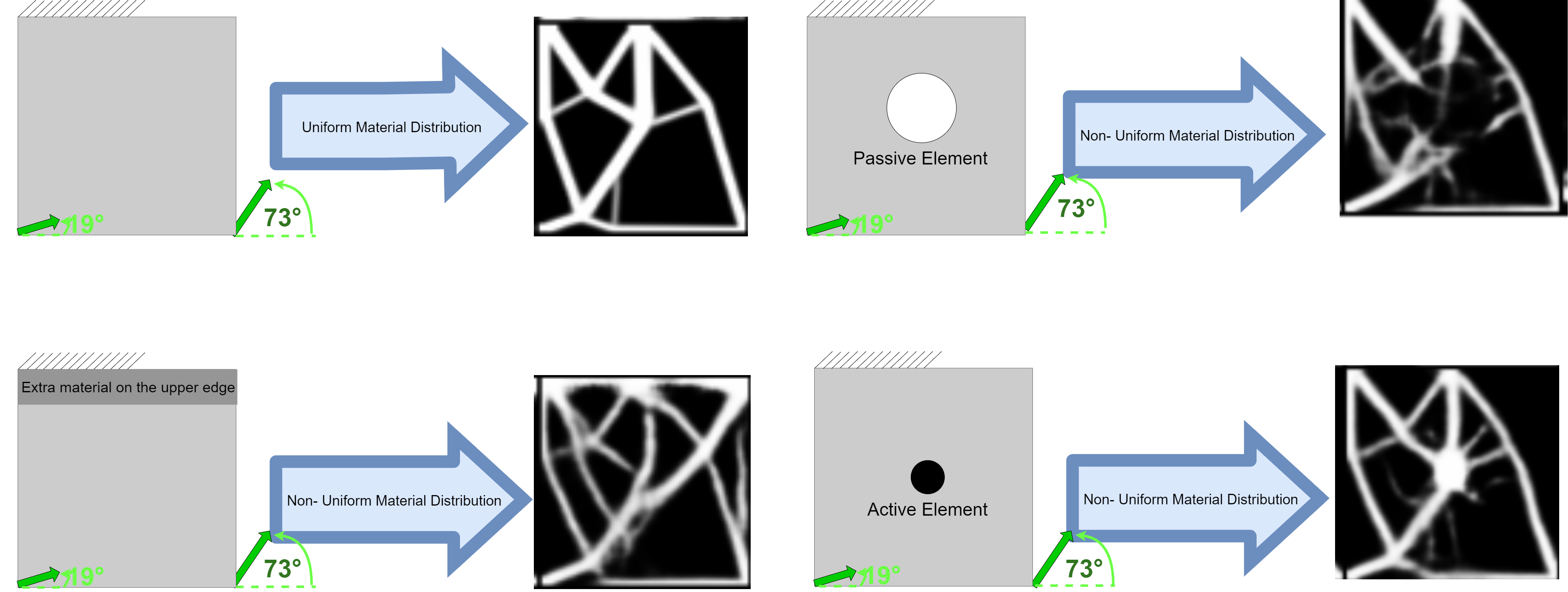}
\caption{GENERATING A DESIGN WITH NON UNIFORM MATERIAL DISTRIBUTION. In this figure, we show the generated design with uniform volume fraction versus those resulting from three types of non-uniform volume fraction.
\label{figure_sketch_non_uniform_volume_fraction} }
\end{figure}
\begin{figure*}[h]
        \includegraphics[width=16cm]{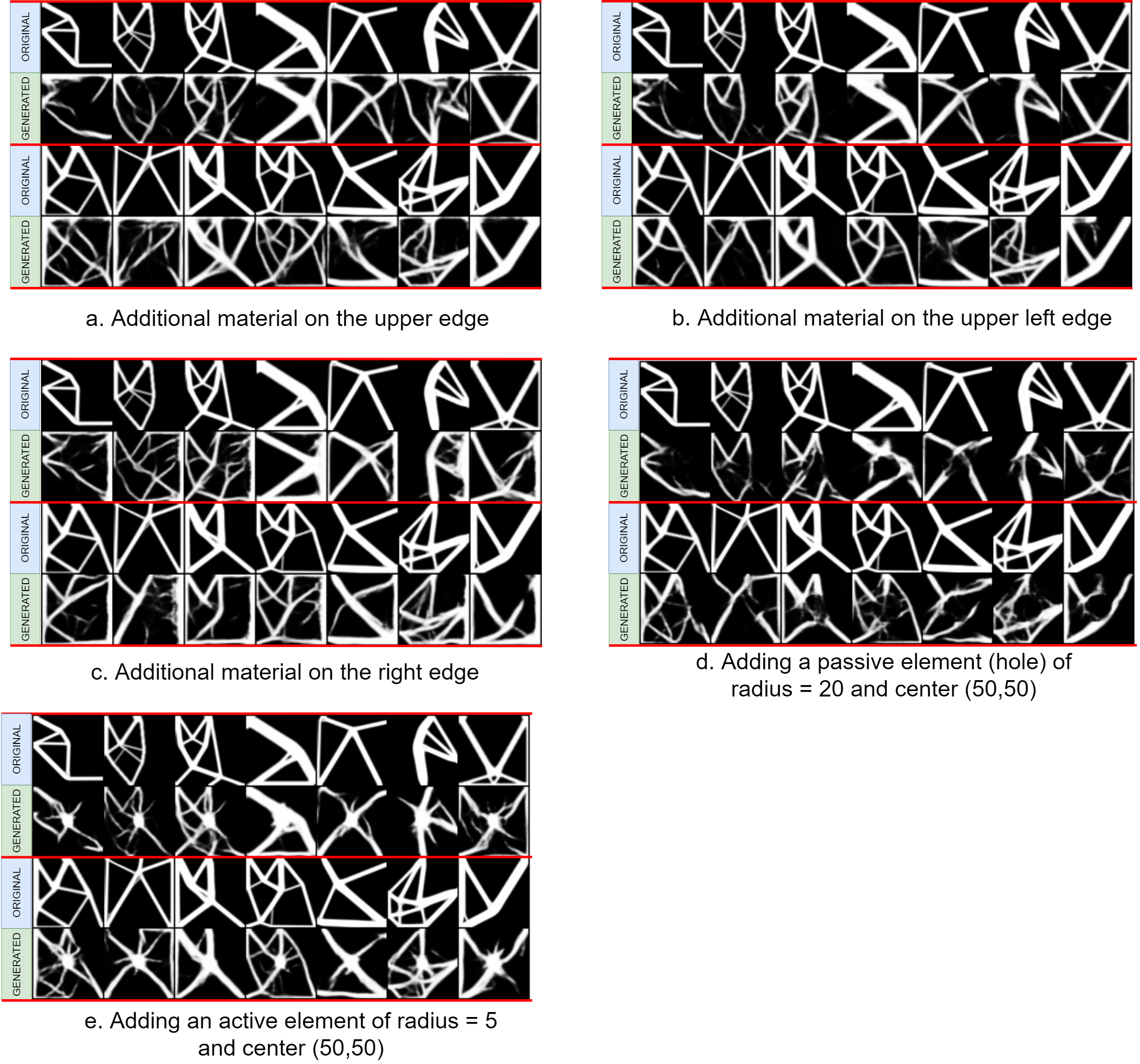}
\caption{DESIGNS WITH NON UNIFORM MATERIAL DISTRIBUTION.
\label{figure_non_uniform_volume_fraction} }
\end{figure*}

\begin{table}[t]
\caption{COMPLEXITY CONSTRAINT. In this table, we show the percentage of generated designs complying with the complexity constraint (i.e. the total number of bars) in the train, validation and test set. \(Cx_g\) and \(Cx_i\) stand for the complexity of the generated designs \(g\) and the input \(i\) complexity constraint (i.e. the complexity of the SIMP-based designs) respectively.}
\label{table_complexity_constraint}
\begin{center}
{\small
\begin{tabular}{ |p{1.5cm}|p{1.5cm}|p{2.5cm}|p{3cm}|}
\hline Dataset & \(Cx_g \leq C_i\) & \(Cx_g \leq Cx_i + 1bar\) & \(Cx_g \leq Cx_i + 2bars\)  \\ 
\hline Train & 89.13\% & 97.94\% & 99.4\%  \\
\hline Validation & 63\% & 73.84\% & 85.4\% \\
\hline Test & 43.3\% & 56.53\% & 67.55\% \\
\hline
\end{tabular}
}
\end{center}
\end{table}

\begin{table}[t]
\caption{VOLUME FRACTION CONSTRAINT. In this table, we show the percentage of generated designs complying with the input volume fraction constraint (i.e. the percentage of material) in the train, validation and test set. \(V_g\) and \(V_i\) stand for the volume fraction computed over the generated designs \(g\) and the input \(i\) volume fraction respectively.}
\label{table_vol_frac_constraint}
\begin{center}
{\small
\begin{tabular}{ |p{1.5cm}|p{1.5cm}|p{2.5cm}|p{2.5cm}|p{2.5cm}|}
\hline Dataset & \(V_g \leq V_i\) & \( \frac{V_g -V_i}{V_i}\%\leq2.5\% \) & \(\frac{V_g -V_i}{V_i}\%\leq5\%\) & \(\frac{V_g -V_i}{V_i}\%\leq10\%\) \\
\hline Train & 60.43\% & 74.33\% & 87.33\%  & 96.78\% \\
\hline Validation & 66.9\% & 79.4\% & 85.18\% & 95.6\% \\
\hline Test & 92.44\% & 96.22\% & 97.79\% & 99.21\% \\
\hline
\end{tabular}
}
\end{center}
\end{table}


\begin{table}[t]
\caption{AVERAGE COMPUTATION TIME OF SIMP (FE-BASED METHOD) VERSUS THE GENERATOR (DEEP-LEARNING-BASED METHOD). In this table, we show the average computational time in seconds (\(s\)) needed to output a 2D design using the SIMP approach vs DL approach.}
\label{table_computation_time_design_generation}
\begin{center}
{\small
\begin{tabular}{ |p{3.5cm}|p{1.2cm}|p{1.2cm}|p{1.2cm}|}
\hline Dataset & Train & Validation & Test \\
\hline SIMP (on CPU) & 642.81 & 140.85 &  206 \\ 
\hline GENERATOR (on CPU) & 0.043 & 0.0389 & 0.0385 \\ 
\hline GENERATOR (on GPU) & 0.0029 & 0.00288 & 0.00289 \\ 
\hline
\end{tabular}
}
\end{center}
\end{table}

As mentioned previously, every design is assessed taking into consideration its conformity with the stress (compliance), volume and complexity constraints. Thus, we evaluate the generator considering structures of varied shapes and number of bars and we calculate for each generated design its compliance, volume fraction and bar count and compare the shape alongside the three metrics listed previously of the generated designs to the original designs outputted by SIMP. The figures Fig.~\ref{figure_validation_set_results_1}, Fig.~\ref{figure_validation_set_results_2} and Fig.~\ref{figure_validation_set_results_3} display the original and the reconstructed designs of the validation set. It is obvious that the shapes of the original (real) designs are well generated, the average reconstruction error (MSE) of the validation designs is 0.055 (Fig~\ref{figure_MSE_distributions}), thus, accordingly the boundary conditions and the load configurations are well respected. Moreover, the tables Tab.~\ref{table_complexity_constraint} and Tab.~\ref{table_vol_frac_constraint} exhibit the percentage of designs in the train and validation sets that comply with the complexity and volume constraints. 85.4\% of the validation designs comply with the complexity constraint with at most 2 extra bars. In other terms, the generated designs tend to have 1 to 2 additional bars at most and this aspect proves that the generator is not memorizing the designs but is being creative. 
In addition, 66.9\% of the generated designs respect the volume constraint and 18.28\% have a volume fraction higher by 5\% than the wanted value, thus, we can conclude that 85.18\% comply with the volume constraint with a relative error margin of 5\% and as Tab.~\ref{table_vol_frac_constraint} shows, 95.6\% comply with the volume constraint with a relative error margin of 10\%. To be more explicit, if the desired input volume fraction is 0.25 i.e. 25\% of material: a 5\% error margin is 26.625\% of material and a 10\% error margin is 27.5\% of material, and hence the difference is negligible. In addition, we have plotted the distributions of the volume fraction and complexity of the generated designs versus the original ones. And it emerges clearly that both the volume fraction as well as the complexity of the generated designs follow a similar distribution of that of the original SIMP-based designs (Fig.~\ref{figure_distributions}). \\
However, the compliance of the generated validation designs seems to be, slightly higher than that of the SIMP-based designs. The average compliance value of the generated validation designs is \(107J\) while the average compliance values of the SIMP-based designs is of \(89J\).
About 67\% of the validation generated designs have a compliance higher by at most 20\% than that of the SIMP-based designs while only 12\% show compliance values that are extremely higher than that of the SIMP-based designs (Fig.~\ref{figure_compliance_distributions}). 
After checking a sample of these extremely high-compliance designs, we have found that some of these designs have a lot of noise pixels (i.e. light grey pixel) and few of them show discontinuities. To overcome this drawback, the generated designs with high compliance can be input to the SIMP-based approach for a few iterations in order to minimize the stress (i.e. the compliance) (Fig.~\ref{figure_designs_high_compliance}).\\
Finally, most of the reconstructed designs comply with the shape, the complexity and volume constraints. However, the compliance values of the generated designs tend to be higher than the original ones. One reason might be that the generator was not penalized explicitly on the compliance during the training. In the future, we can integrate into our GAN-based approach a compliance predictor as a third discriminator so that the generated designs have lower compliance values. We note that the generator was not penalized explicitly on the volume fraction but on the reconstruction and complexity errors, nevertheless, the reconstruction error embeds implicitly the volume fraction constraint. 
In the figure~\ref{figure_validation_set_results_3}, we also show complex designs and their reconstructions. In the case of high frequency detailed designs (i.e. a large number of internal transmission bars), the generator tends to eliminate some internal bars and in some cases to reconstruct them with a very low resolution (blurry).\\
Moreover, we present the  computational advantages of the Deep-Learning-based generator over the SIMP approach (FE-based method) (Tab.~\ref{table_computation_time_design_generation}). The generator is 3500 times (140/0.04) faster in generating a design on CPU and 46000 times (140/0.003) faster on GPU. Even in the rare cases where the generated designs (usually complex ones) showed some discontinuities and where we had to apply a few iterations of SIMP, the generation process is still faster. Thus, using deep-learning, the generation phase is fast enough to allow the designer to explore more designs and improve the mechanical and geometrical properties; knowing that in the industrial world, time is key feature that is not always available for engineers.

To make sure that the generator understands well the notion of complexity, we have fixed the mechanical conditions (BCs, loads and volume fraction) of a sample of designs and played around with the complexity to evaluate the generator's response to such change.
As clearly seen in Fig.~\ref{figure_complexity_variation}, when we increase/decrease the complexity, additional bars appear/disappear in the design while maintaining the volume constraint. However, looking at the compliance value of the reconstructed design versus its counterparts with different number of bars, the compliance always seems to slightly increase for an increase in the complexity while it excessively increases for a decrease of the complexity. We can argue here that a resilient design has a lower bound of number of bars which play an important role in making the design more robust and thus dissipate further the stresses. In fact, some designs, when decreasing their complexity have become discontinuous, thus bringing their compliance values to explode (designs in the third row columns 2 and 7 of Fig.~\ref{figure_complexity_variation}). 

In this work, we contributed not only in integrating the geometrical conditions but also the notion of non-uniform material (density) distribution (i.e. non-uniform volume fraction) into topology optimization; knowing that the generator was trained using only uniform volume fraction. Since, the input volume fraction i.e. the ratio of material is formulated as a matrix of \( (n_x+1\times n_y+1)\) elements, we can always alter this matrix and force the existence/absence of material in specific locations of the design and that by increasing/decreasing the volume fraction values in these particular locations of the matrix. In addition, many geometrical constraints can be integrated using this formulation, for example one can enforce the presence of a passive element (i.e. a hole for a pipe) or an active element (i.e. a filled shape for an external pillar) in the design. We note that the formulations also allow for the addition of boundary and loads conditions over these passive/active elements by simply modifying the BC and loads matrices, however, this aspect was not explored in this research. The types of non-uniform material distribution explored in this study are explained via the sketch in Fig.\ref{figure_sketch_non_uniform_volume_fraction}. We illustrate the results in Fig.~\ref{figure_non_uniform_volume_fraction}; we have used white over black background in order to highlight the modification in the designs. As we can clearly see in the Figure~\ref{figure_non_uniform_volume_fraction}, we chose 5 alterations: adding material to the upper side, the left half of the upper side and the right side, adding a hole of center coordinates \(x_{center}=50, y_{center}=50\) and radius \(r = 20\) and adding an active element of center coordinates \(x_{center}=50, y_{center}=50\) and radius \(r = 5\). In all five cases, the generator filled/emptied the specified locations with material as expected and reshaped creatively the internal truss-like shape to ensure the integrity of the design. As a matter of fact, in most cases, the generated design preserves the outer shape of the original design (i.e. it respects the input boundary conditions and load configurations) and tends to add internal bars and edge bars where we force material in order to ensure the continuity of the design. This promising result is a major contribution to this paper: it goes beyond the traditional topology optimization capabilities in shape and geometry control. Here, we also control the material distribution in the design, a process that cannot be easily accomplished by the traditional approaches of topology optimization. And such achievement gets us closer to our objective in controlling the geometry and complexity of designs with DL in the aim of making the final structures manufacturable. \\
{\it NB: Malvia\cite{malviya2020systematic} has also generated designs with passive and active elements. However, they have trained their generator on such configurations unlike our approach. In our approach, we have not trained the generator on designs including passive or active elements.}

\subsubsection{Test set results - Generalization}

\begin{figure}
\includegraphics[width=16cm]{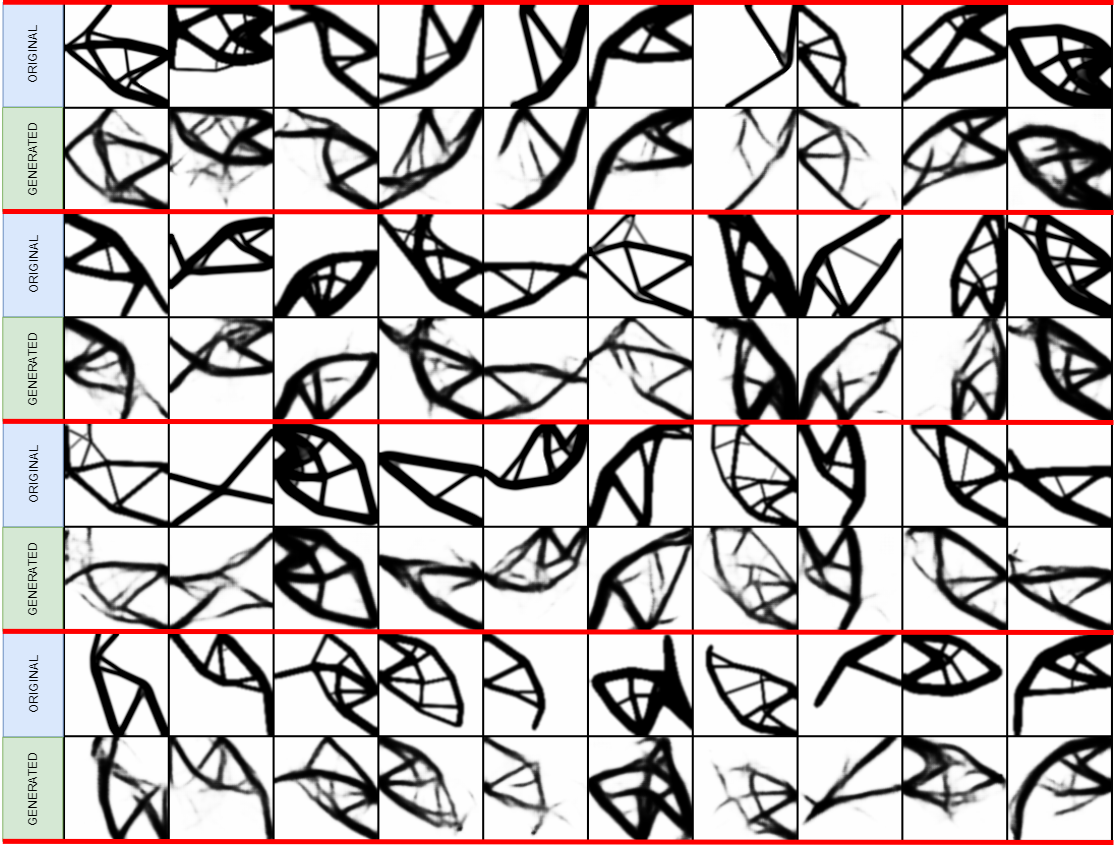}
\caption{ TEST SET RESULTS. These designs were generated considering inner loads.}
\label{figure_test_set_results} 
\end{figure}

In order to test the generalization aspect of the generator, a sample of 635 designs were generated with internal loads. In other terms, the loads distribution is not similar to the one in the train and validation sets. Hence, the data distribution of the test set is different from that of the train and validation sets (Fig.~\ref{figure_distributions} shows the distribution of the complexity and volume fraction of all three sets). The generated designs versus the original ones are shown in the Fig.~\ref{figure_test_set_results}. As clearly seen, the generator is able to reconstruct the outer shape of the designs. However, some finer details (internal transmission bars) in complex designs are omitted or distributed differently in the reconstructed design (the mean reconstruction error is 0.108, i.e. 2 times higher than that of the validation set; Fig~\ref{figure_MSE_distributions}) and this justifies the results shown in Tab.~\ref{table_complexity_constraint}: only 67.5\% of the generated designs comply with the complexity constraint with a maximum of 2 extra bars. As for the volume constraint, it is respected 96.22\% of the time with a 2.5\% error margin i.e. approximately 20\% more than the train/validation's accuracy. One reason would be that this volume fraction precision is a direct consequence of the internal grey bars (i.e. with low density values) present in the generated designs. An additional consequence of the blurry internal bars is the high compliance values seen in the test set: about 75\% of the designs have compliance values that are 2 times higher and more than the ones computed for the SIMP-based designs (Fig.~\ref{figure_compliance_distributions}). As explained in the previous section, few iterations of SIMP can help minimize the compliance (Fig~\ref{figure_designs_high_compliance}).
Despite the compliance drawback, looking at the size of the train set, we can conclude that the generator generalizes very well: it always preserves the outer general shape of the design; in other terms, it respects the boundary conditions; and is creative especially when it comes to the distribution of the internal bars.

\subsubsection{Comparative evaluation }

In this section, we compare our objective evaluation to the evaluations used in the state-of-the-art. As show in the table~\ref{table_evalutation}, Our evaluation is complete and covers all the objectives of this work: the geometrical constraint (complexity), the mechanical constraints (Volume fraction, compliance), the aesthetics (MSE) and the time speedup. Additionally, in this work, automated evaluations were integrated, the complexity and the compliance were computed via DL-based methods.

\section{Conclusion and Future Works}\label{section:conclusion}

This article offers an original novel strategy to accelerate mechanical conception and takes into consideration complex geometrical design criteria using Deep Learning techniques. It is a dual-discriminator GAN approach to generate 2D structures given the mechanical conditions: boundary conditions, loads configurations, volume fraction; and geometrical one: the complexity. It showed great robustness in learning the correlation between both types of conditions and great flexibility in generating creative valid designs. This accelerated process can also be useful to output more robust designs when loads configurations (locations, orientations, magnitudes) and boundary conditions feature uncertainty. Its major contributions apart from the integration of mechanical and geometrical conditions simultaneously into the DL-based topology optimization, are the integration of non-uniform density material into the procedure and the ease of the geometrical condition (i.e. complexity) tuning for fixed mechanical conditions to simplify or refine the structure as a first step towards making it manufacturable. Moreover, this strategy enables, easily, the addition of future developed modules (as discriminators) to the generation approach such as a build time module to identify structures rapidly manufactured, a non-linear material or geometrical effects module, a thermal distortion module and other complex mechanical modules necessary to validate a design mechanically and geometrically. In addition, An objective evaluation has been conducted to ensure the conformity of the designs with the mechanical and geometrical constraints.
In the future, we would like to integrate more complex additive manufacturing conditions and supplementary mechanical validation modules into the generation process and generalize it to 3D structures.

\section*{Acknowledgment}
Thanks go to Dr. Faouzi ADJED for its invaluable technical help and advice provided all along this work.


\bibliography{mybibfile}

\appendix    
\section{}
\label{appendix:A}
\begin{table*}[t]
\caption{GENERATOR RES-U-NET ARCHITECTURE. \(nf\) is the number of feature maps i.e. number of channels.}
\begin{center}
\label{table_GEN_Architecture}
{\small %
\begin{tabular}{|c|c|c|c|c|c|}
\hline & Block Level & Layer & Filter & Stride & Output Size \\
\hline Input &   &   &   &   & \(101\times101\times6\) \\
\hline
Encoder           & Block1 & Downsample (Conv) & \(3\times3/nf\) & 2 & \(50\times50\times nf\) \\ \cline{3-6}
                  &        & Conv1             & \(3\times3/nf\) & 1 & \(50\times50\times nf\) \\ \cline{3-6}
                  &        & Conv2             & \(3\times3/nf\) & 1 & \(50\times50\times nf\) \\ \cline{2-6}
                  & Block2 & Downsample (Conv) & \(4\times4/nf\times2\)  & 2 & \(25\times25\times nf\times2\) \\ \cline{3-6}
                  &        & Conv3             & \(3\times 3/nf\times2\)  & 1 & \(25\times25\times nf\times2\) \\ \cline{3-6}
                  &        & Conv4             & \(3\times 3/nf\times2\)  & 1 & \(25\times25\times nf\times2\) \\ \cline{2-6}
                  & Block3 & Downsample (Conv) & \(3\times3/nf\times4\) & 2 & \(13\times13\times nf\times4\)\\ \cline{3-6}
                  &        & Conv5             & \(3\times 3/nf \times4\) & 1 & \(13\times13\times nf\times4\)\\ \cline{3-6}
                  &        & Conv6             & \(3\times 3/nf\times4\) & 1 & \(13\times13\times nf\times4\)\\ \cline{2-6}
                  & Block4 & Downsample (Conv) & \(3\times 3/nf\times8\) & 2 & \(7\times7\times nf\times8\) \\ \cline{3-6}
                  &        & Conv7             & \(3\times 3/nf\times8\) & 1 & \(7\times7\times nf\times8\) \\ \cline{3-6}
                  &        & Conv8             & \(3\times 3/nf\times8\) & 1 & \(7\times7\times nf\times8\) \\ 
    \hline
Bridge            & Block5 & Downsample (Conv) & \(3\times3/nf\times16\) & 2 & \(4\times4\times nf\times16\) \\ \cline{3-6}
                  &        & Conv9             & \(3\times3/nf\times16\) & 1 & \(4\times4\times nf\times16\) \\ \cline{3-6}
                  &        & Conv10            & \(3\times3/nf\times16\) & 1 & \(4\times4\times nf\times16\) \\ 
    \hline
Decoder           & Block6 & Upsample (TransConv) & \(3\times3/nf\times8\) & 2 & \(7\times7\times nf\times8\) \\ \cline{3-6}
                  &        & Conv11             & \(3\times3/nf\times8\) & 1 & \(7\times7\times nf\times8\) \\ \cline{3-6}
                  &        & Conv12             & \(3\times3/nf\times8\) & 1 & \(7\times7\times nf\times8\) \\ \cline{2-6}
                  & Block7 & Upsample (TransConv) & \(3\times3/nf\times4\) & 2 & \(13\times13\times nf\times4\) \\ \cline{3-6}
                  &        & Conv13             & \(3\times3/nf\times4\) & 1 & \(13\times13\times nf\times4\) \\ \cline{3-6}
                  &        & Conv14             & \(3\times3/nf\times4\) & 1 & \(13\times13\times nf\times4\) \\ \cline{2-6}
                  & Block8 & Upsample (TransConv) & \(3\times3/nf\times2\) & 2 & \(25\times25\times nf\times2\) \\ \cline{3-6}
                  &        & Conv15             & \(3\times3/nf\times2\) & 1 & \(25\times25\times nf\times2\) \\ \cline{3-6}
                  &        & Conv16             & \(3\times3/nf\times2\) & 1 & \(25\times25\times nf\times2\) \\ \cline{2-6}
                  & Block9 & Upsample (TransConv) & \(4\times4/nf\) & 2 & \(50\times50\times nf\) \\ \cline{3-6}
                  &        & Conv17             & \(3\times3/nf\) & 1 & \(50\times50\times nf\) \\ \cline{3-6}
                  &        & Conv18             & \(3\times3/nf\) & 1 & \(50\times50\times nf\) \\ \cline{2-6}
                  & Block10 & Upsample (TransConv) & \(4\times4/\frac{nf}{2}\) & 2 & \(101\times101\times \frac{nf}{2}\) \\ \cline{3-6}
                  &        & Conv19             & \(3\times3/\frac{nf}{2}\) & 1 & \(101\times101\times \frac{nf}{2}\) \\ \cline{3-6}
                  &        & Conv20             & \(3\times3/\frac{nf}{2}\) & 1 & \(101\times101\times \frac{nf}{2}\) \\ \cline{2-6}
                  & Block11 & Conv21             & \(3\times3/1\) & 1 & \(101\times101\times1\) \\ 
\hline
Output &   &   &   &   & \(101\times101\times1\) \\
\hline
\end{tabular}
}
\end{center}
\end{table*}

\begin{figure}
        \includegraphics[width=16cm]{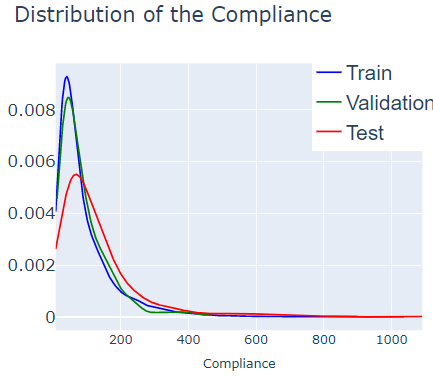}
\caption{DISTRIBUTION OF THE COMPLIANCE VALUES IN THE TRAIN, VALIDATION AND TEST SETS.
\label{figure_compliance_values_distributions_in_sets} }
\end{figure}

\begin{figure}
        \includegraphics[width=13cm, height=20cm]{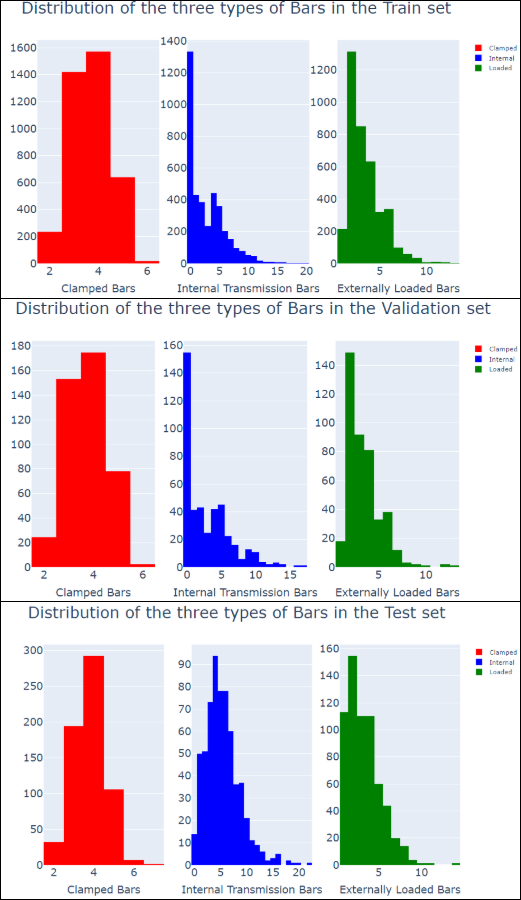}
\caption{DISTRIBUTION OF THE THREE TYPES OF BARS IN THE TRAIN, VALIDATION AND TEST SETS.
\label{figure_type_of_bars_distributions} }
\end{figure}
\begin{figure*}
\includegraphics[width=16cm]{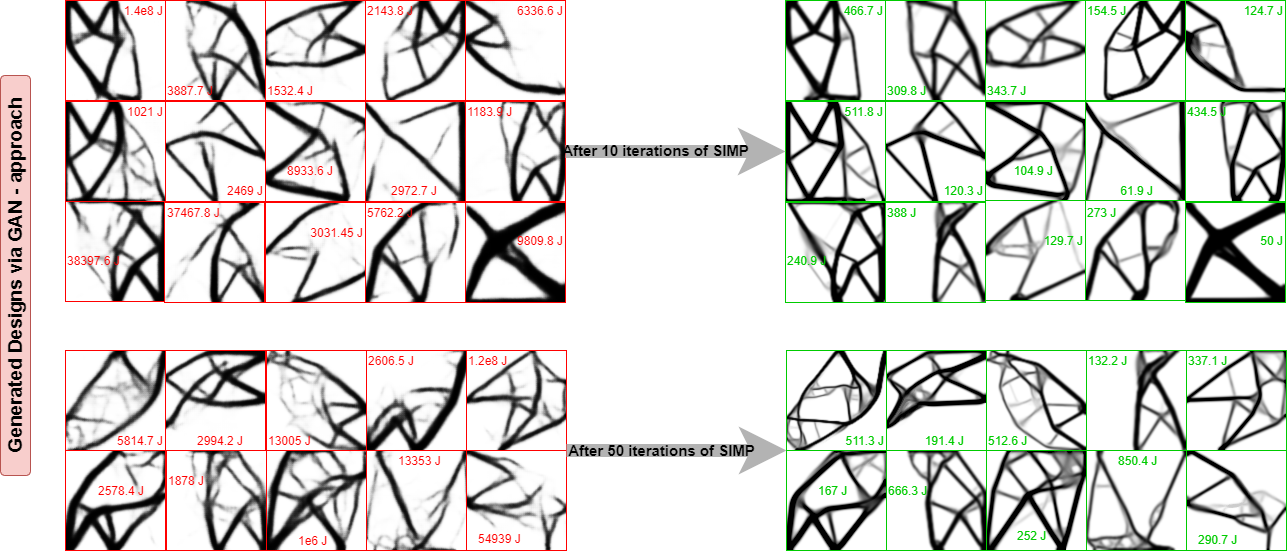}
\caption{GENERATED DESIGNS WITH HIGH COMPLIANCE VALUES. This figure shows a sample of generated design displaying very high compliance that were fed back to SIMP in order to minimize the mechanical stress (i.e. compliance).
\label{figure_designs_high_compliance}}
\end{figure*}

\begin{figure}
        \includegraphics[width=10cm]{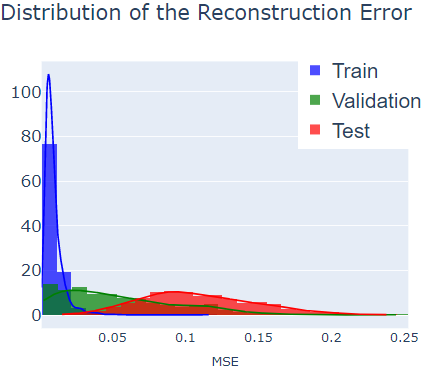}
\caption{DISTRIBUTION OF THE RECONSTRUCTION ERROR IN THE TRAIN, VALIDATION AND TEST SETS.
\label{figure_MSE_distributions}}
\end{figure}

\end{document}